\documentclass[sn-mathphys,Numbered]{sn-jnl}
\usepackage[labelfont=bf]{caption}
\renewcommand{\figurename}{Fig.}
\usepackage{multirow}
\usepackage{nicematrix}
\usepackage{arydshln}
\usepackage{soul}
\usepackage{xcolor}
\usepackage{url}
\usepackage{tikz}
\usetikzlibrary{intersections}
\usepackage{graphicx}
\usepackage{caption}
\usepackage{subcaption, tabularx}
\usepackage{adjustbox}
\usepackage{lscape}
\usepackage{rotating}
\usepackage{float}
\usepackage{multirow}%
\usepackage{amsmath,amssymb,amsfonts}%
\usepackage{amsthm}%
\usepackage{mathrsfs}%
\usepackage[title]{appendix}%
\usepackage{xcolor}%
\usepackage{textcomp}%
\usepackage{manyfoot}%
\usepackage{booktabs}%
\usepackage{algorithm}%
\usepackage{algorithmicx}%
\usepackage{algpseudocode}%
\usepackage{listings}%
\usepackage{float}
\usetikzlibrary{fit}
\usetikzlibrary{arrows,automata}
\usepackage{array, booktabs}
\usetikzlibrary{positioning}
\def\mm#1{\ensuremath{\boldsymbol{#1}}} 

\def\E{\text{E}}
\def\Var{\text{Var}}
\def\Cov{\text{Cov}}
\def\Cor{\text{Cor}}
\def\Diag{\text{Diag}}

\theoremstyle{plain}

\theoremstyle{definition}

\begin{document}

 \title[Graphical Models]{A Graphical Framework for Interpretable Correlation Matrix Models for Multivariate Regression}

\author*[1]{\fnm{Anna} \sur{Freni-Sterrantino}}\email{afrenisterrantino@turing.ac.uk}

\author[2]{\fnm{Denis} \sur{Rustand}}\email{denis.rustand@kaust.edu.sa}

\author[2]{\fnm{Janet} \sur{van Niekerk}}\email{
janet.vanniekerk@kaust.edu.sa}

\author[2]{\fnm{Elias} \sur{Teixeira Krainski}}\email{elias.krainski@kaust.edu.sa}

\author[2]{\fnm{H\aa vard} \sur{Rue}}\email{haavard.rue@kaust.edu.sa}

\affil*[1]{\orgname{The Alan Turing Institute}, \orgaddress{\street{96 Euston Road}, \city{London}, \postcode{NW1 2DB}, \country{United Kingdom}}}

\affil[2]{\orgdiv{Statistics Program, CEMSE Division}, \orgname{King Abdullah University of Science and Technology}, \orgaddress{\city{Thuwal}, \postcode{23955}, \state{ Makkah}, \country{Saudi Arabia}}}

\keywords{Complexity penalized priors,
Correlation Matrix Modelling,
Graphical Structure,
 Multivariate Joint Modelling}

\maketitle

\begin{abstract}

In this work, we present a new approach for constructing models for covariance matrices by considering the decomposition into marginal variances and a correlation matrix, the structure of the latter is deduced from a user-defined graphical structure. The graphical structure makes correlation matrices interpretable and avoids the quadratic increase of parameters as a function of the dimension. We propose an automatic approach to define a prior using a natural sequence of simpler models within the Penalized Complexity framework for the unknown parameters in these models. 

We illustrate this approach with simulation studies of multivariate longitudinal joint modelling, where we demonstrate some properties of the method and two real data applications: a multivariate linear regression of four biomarkers and a multivariate disease mapping. Each application underscores our method's intuitive appeal, signifying a substantial advancement toward a more cohesive and enlightening model that facilitates a meaningful interpretation of correlation matrices.
\end{abstract}

\section{Introduction}

Estimating the covariance matrix and specifying a covariance prior pose challenges from a Bayesian perspective.
 Traditionally, various natural prior conjugate distributions have been used, such as the Wishart~\cite{wishart1928generalised}, inverse-Wishart~\cite{gelmanbda04} and scaled inverse-Wishart~\cite{malley2008domain}. However, the inverse-Wishart prior and its generalization are associated with certain issues. For instance, the uncertainty for all variances is controlled by a single degree of freedom parameter and the marginal distribution for the variances has a low density in a region near zero. There is also a priori dependence between correlations and variances, where larger variances are associated with correlations near $\pm$1, whereas smaller variances are associated with correlations near zero \cite{alvarez2014bayesian}.
In the era of computational frameworks, the attractiveness of preserving prior conjugacy has diminished, as Bayesian inference is mainly conducted using either sampling-based methods like Markov Chain Monte Carlo (MCMC)\cite{Robert1999}, approximate methods such as integrated nested Laplace approximation (INLA)\cite{rue2009approximate} or variational Bayes \cite{blei2017variational}. 

The covariance matrix contains information on two distinct fronts: dependencies (correlations in the correlation matrix) and marginal variances. Thus, a multivariate Gaussian can be reformulated for a vector $\pmb v$ as
\begin{equation}
    \pmb v \sim N_p(\pmb 0, \pmb\Sigma) \quad \rightarrow \quad \pmb v \sim N_p(\pmb 0, \pmb\sigma\pmb C \pmb\sigma^\top).
    \label{eq:intro}
\end{equation}
The correlation matrix is calculated based on the covariance matrix from which the nature and strength of pairwise linear dependence can be extracted. This decomposition allows us to consider priors for the two distinct parts: one for the marginal variances vector and one for the correlation matrix.
For the latter, two major challenges emerge (i) the unit diagonal and positive definite constraints; (ii) the number of parameters grows quadratically as the
number of variables increases. The typical approach assumes correlation among variables (or random effects), but these are not explicitly elicited. Hence, how variables are correlated is unknown and unaccounted for in the prior choice.

Proposals for correlation priors include priors based on the spectral decomposition of a symmetric matrix \cite{Jin2007, MacNab2018} and priors on the singular value decomposition of an asymmetric matrix \cite{Greco2009,macnab2016linear}. Both are characterized by excessive informative constraints on the correlation matrix elements, leading to shrinkage estimation towards the matrix diagonal.
Another proposal is the LKJ prior \cite{lewandowski2009generating}, which is a weakly informative prior defined on a simplex. Tuning the strength of just one parameter allows to control how closely the sampled matrices resemble the identity matrices.

Nonetheless, these proposals' priors tend to shrink the estimation of the correlation matrix towards the identity matrix. This feature can be problematic as it excludes models that have varying degrees of correlation and are not limited to being either fully correlated or fully uncorrelated. 
Therefore, there is a need for more interpretable models that can account for correlation matrices with different structures, separated from the marginal variances.

We develop a framework to derive a prior for the correlation matrix $\pmb C$ in \eqref{eq:intro} (thus inherently for the covariance matrix $\pmb\Sigma$, as well), based on a graph of dependence. This approach addresses two main challenges. Firstly, we reparameterize the correlation matrix based on a latent dependence graph to reduce the dimension of the estimable parameters of the correlation matrix. Secondly, we develop an intuitive prior that shrinks the dependence quantified in the correlation matrix to a sequence of simpler dependence models.

A latent model is deduced to control the quadratic growth of parameters for high-dimensional models, as the correlation matrix $\pmb C$ tends to become large and can involve many correlations to be estimated. Thus, we can explain some of the correlations between response variables through their relationship to latent factors. One class of latent models is the multivariate Gaussian Markov random field (MGMRF) model.
These models find their most popular application in spatial statistics, i.e. in multivariate disease mapping, using multivariate conditional autoregressive model \cite{MacNab2018,Jin2005}; in 
longitudinal and survival analysis \cite{Rustand23, Rustand23_2}, spatio-temporal datasets \cite{vicente2020bayesian,macnab2022bayesian}, and describing nature's and systems' multivariate spatial and temporal dynamics \cite{Rue2005,boaz2019multivariate,lee2017multivariate}. Each component of the MGMRF is a GMRF with possible random effects that can have the usual dependence across space, time, units etc. In this work, we propose a method on how to link the various components in an MGMRF in an intuitive way based on a user-constructed graph.
Based on the elicited correlation structure between GMRF's from the graphical construction, we adopt the penalizing complexity (PC) prior (by definition weakly informative)\cite{Simpson2017} framework to develop a prior based on a sequence of simpler correlation matrices models. This prior contracts toward a simpler model (base model), which is not necessarily the identity matrix. 
We implemented this approach within the latent Gaussian process in \textbf{R-INLA}, allowing for low-dimensional parameters. Hence, this approach has been developed for multivariate response modelling when the dimension is manageable ($\leq 10-15$).  

The remainder of this paper is the following: in Section \ref{sec:comp}, we show how to derive correlation matrices from a graphical construction. Then, in Section \ref{sec:base}, we describe how we derive a sequence of simpler models for correlation matrices using the graphical framework. We present the penalized complexity priors principles in Section \ref{sec:pcom}, and we define the PC prior for a correlation matrix. In Sections \ref{sec:sim} and \ref{sec:data}, using the proposed modelling approach, we provide a simulation study on longitudinal joint models and two real data applications of multivariate analysis, respectively. Concluding remarks are given in Section \ref{sec:disc}.

\section{Graph-derived correlation matrices} 
\label{sec:comp}

\subsection{Dependence deduced from a graph} 

Correlation is interpreted as the association between two variables by quantification of the degree of joint variability. Often two variables are correlated because they share a common factor.
This perspective suggests that the observed correlation between two variables may arise from both variables being influenced by latent or unobservable factors. Within this theoretical framework, the observed correlation does not necessarily signify a direct causal relationship between the variables; rather, it underscores their collective responsiveness to shared influences. 

To summarize the correlation between variables, the graphical Gaussian models (GGM) \cite[chp. 7] {KollerDaphne} offer a visual and computational framework to organize the conditional independence in sparse matrices. GGMs focus on modelling the relationships between observed variables directly. Each observed variable corresponds to a node in the graphical representation, and edges in the graph represent conditional dependencies between variables. On the other hand, the absence of an edge between them implies that they are conditionally independent, given all other variables in the model, corresponding to a zero entry in the precision matrix 

However, if the correlation is constructed as two variables sharing a common effect, GGMs are extended to include latent (unobserved) variables. These latent variables capture shared information among the observed variables, allowing the model to represent more complex relationships. Latent GGMs provide greater flexibility in capturing dependencies, 
allowing for a more comprehensive representation of the underlying structure in the data. By capturing heterogeneity in the data by allowing for the presence of unobserved factors to contribute to variability. This is particularly useful in scenarios where subgroups or hidden patterns influence the relationships among variables.

Based on the Latent GGM, we define a tree graph where the correlations between two nodes are given by sharing a latent ancestor node, which will induce the children's correlation. Edges connect only a latent node to children or to other latent nodes, with no edges between children and only one edge to the latent parent, with a rooted node as a latent factor. 
From the graphical representation, we factorize the joint in terms of local conditional probabilities. This provides a sparse matrix for the precision matrix, with a dense block for the observed (children) nodes. 
While sparse matrices come with their computational benefits, the dense matrix obtained by this graphical structure leverages the advantages of latent GGMs and offers an interpretable framework for understanding the intricate relationships within the data. The inclusion of the parent latent nodes avoids specifying directly the correlation among children and guarantees a valid correlation matrix.

For interpretability and illustration purposes, we show how we can derive the correlation matrices from graphs with three latent factors in the next section.

\subsection{Illustrative example (A correlation matrix with three parameters)}
\label{sec:sub23}

\begin{figure}[h!]
     \centering
    \begin{tikzpicture}[shorten >=1pt,node distance=2.5cm,on grid,auto]
    \tikzstyle{state}=[shape=circle,thick,draw,minimum size=0.6cm]
  \tikzstyle{statesq}=[shape=rectangle,thick,draw,minimum size=0.6cm]
    \node[statesq,fill={rgb:black,1;white,2}] (c1) {$p_1$};
    \node[statesq,fill={rgb:black,1;white,2}, below left of=c1, yshift=5mm] (b1) {$p_2$};
    \node[statesq,fill={rgb:black,1;white,2}, below right of=c1, yshift=5mm] (b2) {$p_3$};
    \node[state, below of=b2] (Z4) {$c^*_4$};
    \node[state,below left of=b1] (Z1) {$c^*_1$};
    \node[state, below of=b1] (Z2) {$c^*_2$};
    \node[state, below right of=b1] (Z3) {$c^*_3$};

    \path[->,draw,thick]
    (c1) edge node {} (b1)
    (c1) edge node {} (b2)
    (b2) edge node {} (Z4)
    (b1) edge node {} (Z1)
    (b1) edge node {} (Z2)
    (b1) edge node {} (Z3) ;
  \end{tikzpicture}
        \caption{Three parameters graph with three variables. Grey and white nodes represent latent factors and observed variables, respectively.}
        \label{fig:fig3}
\end{figure}
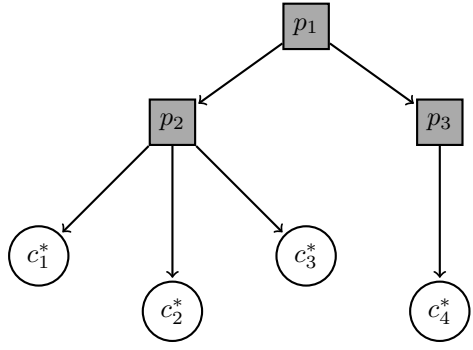
Suppose we have four variables (observed or latent), $\{c_1,c_2,c_3,c_4\}$. We formulate a joint Gaussian density for them such that  $\pmb c \sim N_4(\pmb 0, \pmb\sigma \pmb C\pmb\sigma^\top)$, where $\pmb\sigma$ is the vector of marginal variances of $\pmb c$.

Given the (directed) graph in Figure \ref{fig:fig3}, we assume that the scaled children nodes $c^*_i = \sigma_i^{-1}c_i, i=1,\dots,4$ represent the variables under consideration that are dependent through three parent latent factors $p_i,i=1,..,3$. 
The latent factor $p_1$ induces correlation between the four variables $c_i$ with $i=1,..,4$ and the two latent factors $p_2$ and $p_3$. We consider the conditional distributions:
\[
\begin{matrix}
c^*_i|p_2 \stackrel{\text{iid}}{\sim} \mathcal{N}(p_2,1), i=1,2,3\\
c^*_4|p_3 \sim \mathcal{N}(p_3,1)\\
p_1 \sim \mathcal{N}(0, q_1^2)\\
p_2|p_1\sim \mathcal{N}(p_1,q_2^2)\\
p_3|p_1 \sim \mathcal{N}(p_1,q_3^2)\\
\end{matrix}
\]

It is easy to construct the correlation matrix from the full joint (Gaussian)
distribution. As the $c^*_i$ are independent given $p_i$, the joint distribution for this Gaussian
Markov Random Field (GMRF)[15] for fixed ($q_1,q_2,q_3$) is: 
\[
\pi(c^*_1,c^*_2,c^*_3,c^*_4,p_2,p_3,p_1|q_1,q_2,q_3)\propto\\ \exp\left (-\frac{1}{2} \left (\frac{p_1^2}{q_1^2} +\sum_{i=1}^{p=3}(c^*_i-p_2)^2+(c^*_4-p_3)^2+ 
\frac{(p_2-p_1)^2}{q_2^2} + \frac{(p_3-p_1)^2}{q_3^2}\right) \right) 
\]

To obtain the precision matrix for this GMRF, we need to compute the Hessian of the negative logarithm of this distribution. The Hessian matrix carries the conditional information of the random vector $\mm{\theta}=(c^*_1,c^*_2,c^*_3,c^*_4,p_1,p_2, p_3)$, when all the other parameters are fixed. This also follows by the definition of a GMRF (Theorem 2.2 in \cite{Rue2005}) where the diagonal elements of the precision matrix express the conditional precision given the other variables; the off-diagonal elements correspond to pairwise correlation between the nodes of a graph, conditional on the other nodes.

It follows that the precision matrix is: 
\begin{equation*}\label{equsub2}%
    \mm{\Sigma}^{-1}=\begin{bmatrix}{} 
          1 & . & . & . & -1 &.      & .  \\ 
         . & 1 & . & . & -1 & .      & .  \\ 
         . & . & 1 & . & -1 &       &.  \\
         . & . & . & 1 & .   & -1  & . \\
         -1 & -1 &-1 & . & 3+1/q_2^2 & .&-1/q_2^2 \\
        .  & . &. &-1 &. & 1+1/q_3^2 &-1/q_3^2 \\
        .  & . & . & .  & -1/q_2^2& -1/q_3^2 & 1/q_1^2+1/q_2^2+1/q_3^2\\
              \end{bmatrix}.
\end{equation*}

Zero elements are indicated as '$.$'. Finally, we extract the left upper block of dimension $ 4 \times 4$ of $C$ (i.e., the number of child nodes):
\begin{equation}
\pmb{C}=\Diag(\mm{\Sigma})^{-1/2} \mm{\Sigma}~\Diag(\mm{\Sigma})^{-1/2}.
\end{equation}

\begin{equation*}\label{eqc2}%
   \pmb C[1:4,1:4]=\begin{bmatrix}{} 
         1        &  \rho_{1}  & \rho_{1}    &  \rho_{2}   \\ 
         \rho_{1}  & 1 &  \rho_{1}     &  \rho_{2}    \\ 
         \rho_{1}  & \rho_{1}    & 1  &  \rho_{2}  \\
          \rho_{2} &  \rho_{2}  &   \rho_{2}  &  1 \\
    \end{bmatrix}
\end{equation*} 
where
\[
  \rho_1=\frac{q_1^2+q_2^2}{q_1^2+q_2^2+1}
\]
 and 
\begin{equation}
    \label{eqrho2}
    \rho_2=\frac{q_1^2}{\sqrt{q_1^2+q_2^2+1}\sqrt{q_1^2+q_3^2+1}}
\end{equation}
where $\rho_1$ indicate the correlation between $c^*_1$ and $(c^*_2,c^*_3)$ and $\rho_2$ indicate correlation between $c^*_4$ and $(c^*_1,c^*_2,c^*_3)$. This example illustrates the fact that instead of $6$ distinct correlation parameters we now only have $2$ correlations.
In the case of three latent nodes, we observed that the correlation among $(c^*_1,c^*_2,c^*_3)$ is given by $\rho_1=f_1(q_1, q_2)$, while $\rho_2=f_2(q_1,q_2,q_3)$. The latter correlation also accounts for some variability that is in $p_2$. 

Then, the Gaussian prior for $\pmb c
 = \{c_1,c_2,c_3,c_4\}$ can be formulated as:
 \begin{equation}
     \pmb c \sim N_4(\pmb 0, \pmb\sigma \pmb C\pmb\sigma^\top),
 \end{equation}
 with the hyperpriors
\begin{eqnarray}
     \sigma_i &\sim& \pi(\sigma_i), \quad i=1,2,3,4\\ \notag
     \pmb C(\rho_1, \rho_2) &\sim& \pi(\rho_1,\rho_2) = \pi(q_1,q_2,q_3).
\end{eqnarray}
The correlation matrix $\pmb C$ depends on only three parameters, and the matrix-variate prior can thus be formulated as an interpretable trivariate prior.\\

Extending these models to more variables is straightforward,
It is easy to see that it is sufficient to express the joint distribution of the graph models and that the derived correlation matrix is a function of the latent nodes' variances.

\subsection{An explicit result}

Using the law of total expectation, variance and covariance, we can directly compute $\Cor(c_i, c_j)$, $i\not=j$ from the graph. As an example let us compute $\rho_2 = \Cor(c^*_1, c^*_4)$ in Figure~\ref{fig:fig3}. We get that $\E(c^*_i)=0$, $\Var(c^*_1) = 1 + q_2^2 + q_1^2$, $\Var(c^*_4) = 1 + q_3^2 + q_1^2$ simply by following the reverse paths in the graph from $c^*_1$ and $c^*_4$ to $p_1$. For the covariance, then $\Cov(c^*_1, c^*_4) = q_1^2$, as $p_1$ is the first parent node where the reverse paths from $c^*_1$ and $c^*_4$ to $p_1$ meet, and $\Var(p_1) = q_1^2$. These results gives $\rho_2$ in equation~\ref{eqrho2}.\\ \\
Note that we will subsequently use the notation $\pmb c$ instead of differentiating between $\pmb c$ and $\pmb c^*$ (although the premise remains the same),  where the graphs inform only the conditional dependency between the elements of $\pmb c$ and not the scale, as in this section.

\section{Constructing a sequence of simpler correlation matrices}
\label{sec:base}

\begin{sidewaysfigure}[ht]
\centering
        \includegraphics[scale=0.7]{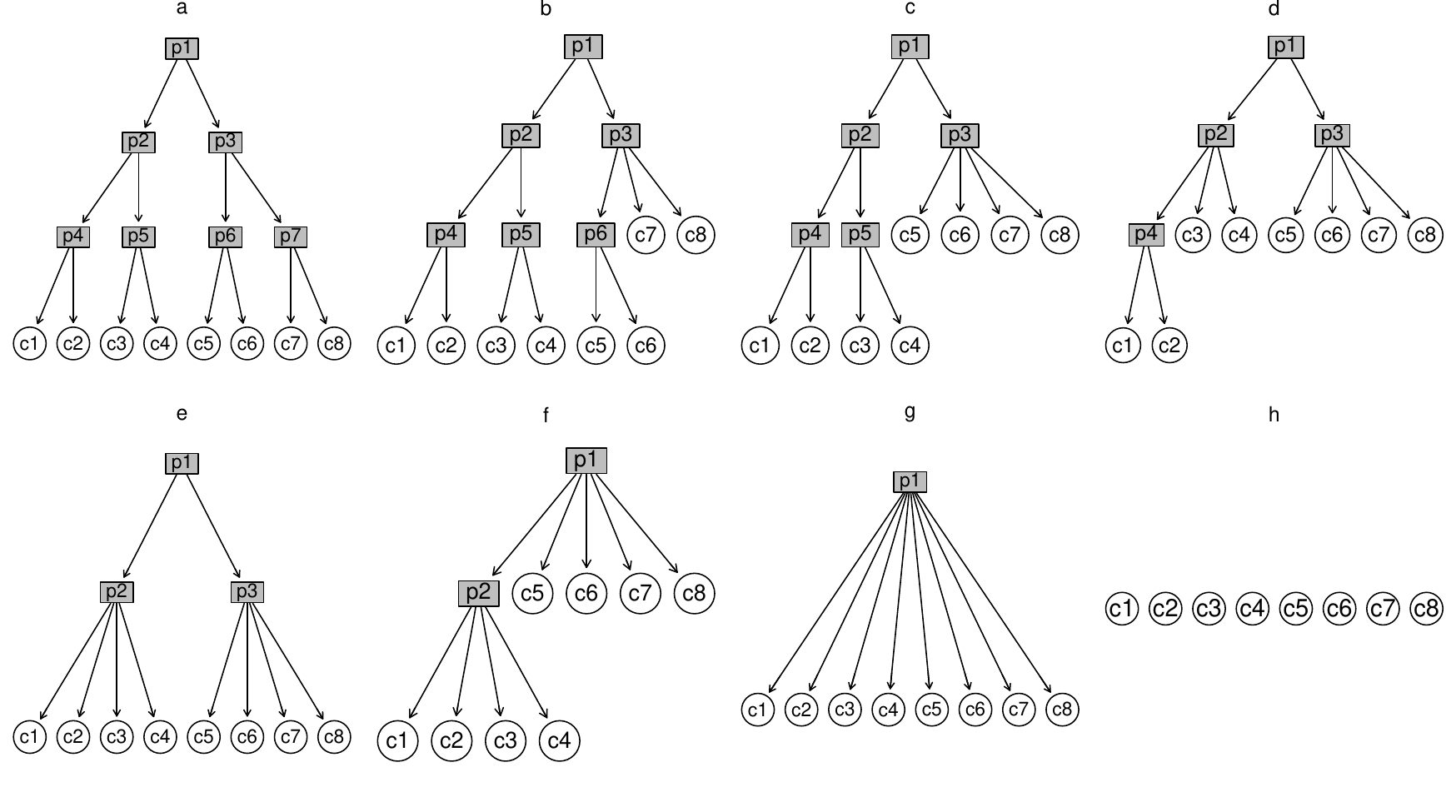}
      \caption{Graph sequence that generates correlation matrices, from the complex $a$ to the simplest $h$ (identity correlation matrix), for 8 observed variables or random effects (circle nodes) and 7 latent nodes (square grey nodes).}
        \label{fig:basemod}
\end{sidewaysfigure}

The correlation coefficients are functions of the parent's variances as shown in Section \ref{sec:sub23}. By further exploiting the graph, we can automatically generate a sequence of simpler correlation matrices by removing one parent at a time. Based on this approach, there are as many simpler models as there are parent nodes that will represent parsimonious models.
It is clear why the identity matrix could be a reasonable choice only when the correlation matrix is an exchangeable matrix. When the correlation matrix is unstructured, choosing an uncorrelated as a simpler model might cause too much shrinkage.  In Figure \ref{fig:basemod}$a$, we show an example of how, from the graph with eight children and seven parents, by removing, in order, the parent node, we generate a sequence of seven graphs, with the last two devising an exchangeable (Figure \ref{fig:basemod}$g$) and an uncorrelated matrix (Figure \ref{fig:basemod}$h$). 

It is important to note there is a correlation magnitude that should be expected in the different steps. 
For example, we can compare the correlation between three sets of children between graphs $e$, the complex model, and $f,g$, the simpler models, from Figure \ref{fig:basemod}. The correlation equations are the following: 

\renewcommand{\arraystretch}{2}
  \begin{table}[h]
        \centering
\large
\begin{tabular}{ c c c}
$\overbrace{\rule{4.5cm}{0pt}}^{\text{Graph } e} $  & $\overbrace{\rule{4.5cm}{0pt}}^{\text{Graph } f~(q_3^2=0)}$   & $\overbrace{\rule{4.5cm}{0pt}}^{\text{Graph } g~(q_2^2=0)} $ \\  
            $\rho_{children 12}= \frac{q_1^2+q_2^2}{1+q_1^2+q_2^2} $ & $\rho_{children 12}= \frac{q_1^2+q_2^2}{1+q_1^2+q_2^2} $  &   \\  
           $\rho_{children 15} = \frac{q_1^2}{\sqrt{1+q_1^2+q_2^2}\sqrt{1+q_1^2+q_3^2}} $ & $\rho_{children 15} = \frac{q_1^2}{\sqrt{1+q_1^2+q_2^2}\sqrt{1+q_1^2}}$   &     $\rho_{Exchangeable}=\frac{q_1^2}{q_1^2+1}$\\
            $\rho_{children 56} = \frac{q_1^2+q_3^2}{1+q_1^2+q_3^2}$ & $\rho_{children 56} = \frac{q_1^2}{1+q_1^2}$ &  
        \end{tabular}
    \end{table}
\noindent where $\rho_{children 56}$, from graph $f$ equals $\rho_{Exchangeable}$.

In the complex model panels (graph $e$), correlation $\rho_{children 12}$ and $\rho_{children 56}$ dominate over $\rho_{children 15}$.

In the simpler model panels (graph $f$), when $q_3^2=0$, $\rho_{children 12}$ continues to dominate over $\rho_{children 15}$, while the correlation $\rho_{children 56}$ depends only by $q_1^2$ , and increases to match the correlation with $\rho_{children 12}$ for higher values of $q_1^2$. In the graph denoted as $g$, correlations exhibit uniformity and increase with higher values of $q_1^2$. The last simpler model is always the identity matrix for $q_1^2=0$.
This result provides a way to assess how the variables (children nodes) can be grouped and what correlation magnitudes should be expected by the graph construction; hence, it can be reorganised to account for that. In more sophisticated models, when the graphs describe the correlation between random slope and intercepts, it helps to grasp the correlation structure.
In summary, the parent removal process gradually contracts toward the uncorrelated matrix. This insight serves as a foundation for constructing a prior for these model components within the framework of a penalized complexity prior, which are going to be introduced next.

\clearpage
\clearpage

\section{Penalized complexity prior for \texorpdfstring{$\pmb C$}{}}
\label{sec:pcom}

The core of the framework relies on specifying a base (or simple) model, that represents the simple version and the complex or flexible model that can be viewed as an extension of the parsimonious base model. 
For example, the skew-normal is an extension of the normal model, a random effects model is an extension of a fixed effects model, a Weibull model is an extension of the exponential model and so on. Penalized complexity priors (PC-priors) are a class of weakly informative priors that penalize departures from the complex model to the base model, as introduced by Simpson et al. (\cite{Simpson2017}).
The core idea is that for a model we can identify a `flexibility' parameter $\xi$. 
Suppose $f_1$ is a simple Gaussian model with an iid random effect. We need to specify a prior on the precision of the random effect ($\tau$). The base model density $f_0$ corresponds to a fixed
value of the parameter $\tau =0$ (i.e. the flexibility parameter), hence a model with no random effect.
If the model has flexibility parameters, we set a prior on it that penalises the complexity of the model. With a prior mass to be near the base value. Each component of a model has a prior penalising the complexity.

The concept of PC priors has been extend to a wide range of modelling approaches such as: autoregressive modelling \cite{sorbye2017penalised}, spatially correlated areal data \cite{Riebler2016}, for the range and marginal variance parameters of a Gaussian Random Field \cite{fuglstad2019constructing} and varying coefficient models \cite{franco2019unified}.

\subsection{Preliminaries}\label{sec:pcprelim}
Briefly, we present the four principles of a PC prior:

\begin{enumerate}
\item \emph{The Occam's razor parsimony}. Unless there is evidence to move toward a complex model, the prior should penalise departure from the base model. 
  
\item \emph{Measure of complexity}. To compute the measure of complexity between the base and complex model, the Kullback-Leibler Divergence (KLD) 
  \[
KLD(\pi(\mm{x}|\xi)|| \pi(\mm{x}|\xi=0))=\int \pi(\mm{x}|\xi)\log\left(\frac{\pi(\mm{x}|\xi)}{ \pi(\mm{x}|\xi=0)}\right)\emph{d}\mm{x},
 \]
 is used. The $\xi$ is the flexibility parameter, for which we define the prior. 
  The KLD measures the information lost when the complex model is approximated with the base model. For two normal distribution $f_0$ and $f_1$  where
 $f_i \sim \mathcal{N}(0,\mm{\Sigma}_i)$ $i = 0,1$, while $n$ is the covariance matrix dimension , the KLD is:
 \begin{equation*}
  \label{eq:kld}
KLD(f_1 || f_0)=\frac{1}{2}\bigl\{tr(\mm{\Sigma}_0^{-1} \mm{\Sigma}_1)-n-\log \left( \frac{|\mm{\Sigma}_1|}{|\mm{\Sigma}_0|}\right)\bigl\}.
\end{equation*}
The KLD is transformed to a unidirectional distance measure,
\[
  d(f_1 || f_0)=\sqrt{2KLD(f_1 || f_0)}.
\]
By Pinsker's inequality, this is also an upper bound of the total variation distance between $f_0$ and $f_1$. The prior is then defined as a function of the distance $\pi(d)$, with a high mass on areas where replacing the flexible model with the base model will not compromise too much loss of information, i.e. where $d=0$. 

A prior that contracts to $d=0$ avoids overfitting because it allows the base model to emerge in the posterior distribution. In contrast, priors that approach zero at $d=0$ can lead to overfitting by pulling the posterior away from the base model, even if the base model is the true one.

\item \emph{Constant rate of penalisation}. The prior distance should be chosen in a way that the mode should be located at the base model, while the density decays as the distance from the base model
increases. We chose a constant rate of penalization that satisfies the following:
  \[
\frac{\pi_d(d+\delta)}{\pi_d(d)}=\frac{\lambda \exp(-\lambda (d+\delta))}{\lambda \exp(-\lambda d)}=r^{\delta},
  \]
with $\delta, d \geq 0$ and $0<r<1$, where where $r=\exp(-\lambda)$ is the constant decay rate.  This means that the prior is independent of the actual distance. The resulting is an exponential prior is formulated for the distance 
\[
\pi_d(d)=\lambda \exp(-\lambda d),
\]
 a change of variable in the prior for the flexibility parameter $\xi$, returns:
 \begin{equation*}
  \label{eq:prior}
\pi(\xi)=\lambda \exp(-\lambda d(\xi)) \left|\frac{\partial d(\xi)}{\partial \xi}\right|.
 \end{equation*}

\item  \emph{User-defined scaling}. The PC-prior has one user-specified rate parameter: $\lambda$. This parameter controls the probability mass at the tail, and the user can thus intuitively specify this value using the following formulation:
\[
Prob(\mathcal{Q}(\xi) > U)=\alpha,
\]
Where $\mathcal{Q}(\xi)$ is a transformation of the flexibility parameter into an interpretable quantity, as the distance itself is typically not directly interpretable. 
The prior information can then be included, for example, tail probabilities $P(Q(\xi) > U )= \alpha$ or $P(Q(\xi) < L)= \alpha$, where U and L are the 
an upper or lower limit, respectively, and $\alpha$ is the upper or lower tail probability of the
prior distribution.
$L/U$ are a user-defined bounds for the tail event, and $\alpha$ is the weight assigned to the event. From the properties of the exponential distribution, we can see that we can satisfy this if we choose
$\lambda=\alpha/{d^{-1}(Q^{-1}(U))}$ \cite{simpson2022}.
\end{enumerate}

The penalized complexity prior is a tilted Jeffrey's prior for $\xi$. We may derive it by leveraging the relationship between the Kullback-Leibler divergence and Fisher information computed at the base model for exponential families \cite{Simpson2016}. For each model couple in the sequence, we can approximate the KLD as: 
\begin{equation}
    \label{eq:eqkld}
  KLD(\pi(\mm{x}|\xi)|| \pi(\mm{x}|\xi=0))=\frac{1}{2}I(0)\xi^2 +\text{higher order terms} ,
\end{equation}
where $I(0)$ is the Fisher information computed at the base model where $\xi=0$, represents the variance associated with the parent removed. It is easy to verify that the prior is then:
\[
\pi(\xi)=\lambda \sqrt{I(0)} \exp(-\lambda \sqrt{I(0)}\xi).
\]
The probabilities of the exponential distribution are tilted with respect to $\sqrt{(I(0)}$. However, in our code, we have implemented the exact numerical version, see Appendix \ref{appendix:appA}.

In summary, the PC prior is defined on the scale of the distance from the base model and then transferred to the scale of the original parameter by a standard
change of variable transformation. We can build the prior for the corresponding flexibility parameter using an exponential distribution based on the distance from the base model and then transform it back to the original scale. In the end, the suitable prior is the one place most of its mass on $d=0$, and this requirement helps to guide the choice of the $\lambda$, which will depend on the change of variable on the flexibility parameter.

\subsection{Penalized complexity prior for the graph-based correlation matrix}

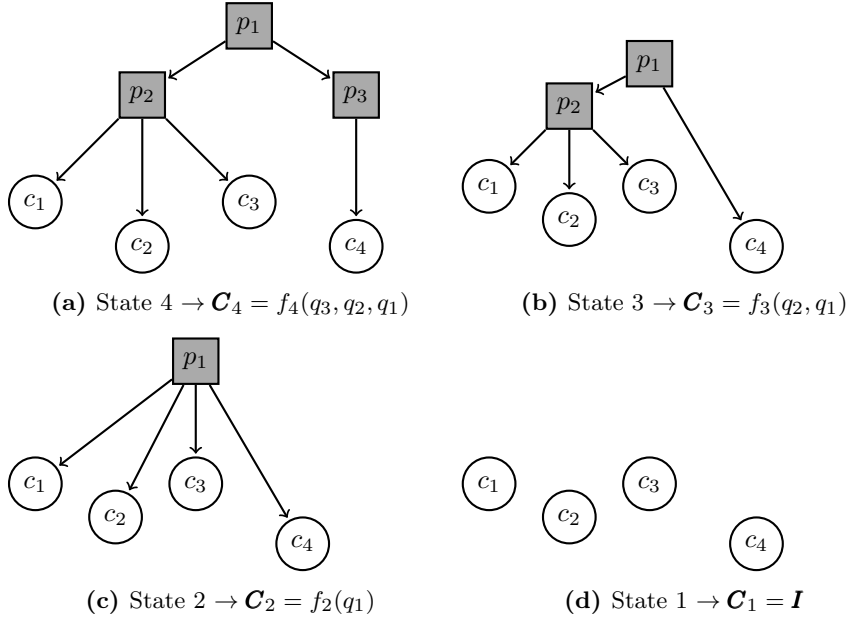
\begin{figure}[h!]
     \centering
     \begin{subfigure}[b]{0.45\textwidth}
      
    \begin{tikzpicture}[shorten >=1pt,node distance=2cm,on grid,auto]
    \tikzstyle{state}=[shape=circle,thick,draw,minimum size=0.6cm]
  \tikzstyle{statesq}=[shape=rectangle,thick,draw,minimum size=0.6cm]
    \node[statesq,fill={rgb:black,1;white,2}] (c1) {$p_1$};
    \node[statesq,fill={rgb:black,1;white,2}, below left of=c1, yshift=5mm] (b1) {$p_2$};
    \node[statesq,fill={rgb:black,1;white,2}, below right of=c1, yshift=5mm] (b2) {$p_3$};
    \node[state, below of=b2] (Z4) {$c_4$};
    \node[state,below left of=b1] (Z1) {$c_1$};
    \node[state, below of=b1] (Z2) {$c_2$};
    \node[state, below right of=b1] (Z3) {$c_3$};

    \path[->,draw,thick]
    (c1) edge node {} (b1)
    (c1) edge node {} (b2)
    (b2) edge node {} (Z4)
    (b1) edge node {} (Z1)
    (b1) edge node {} (Z2)
    (b1) edge node {} (Z3) ;

  \end{tikzpicture}
  \caption{State 4 $\rightarrow \pmb C_4 = f_4(q_3,q_2,q_1)$}
\end{subfigure}
 \begin{subfigure}[b]{0.45\textwidth}
   \begin{tikzpicture}[shorten >=1pt,node distance=1.5cm,on grid,auto]
    \tikzstyle{state}=[shape=circle,thick,draw,minimum size=0.6cm]
  \tikzstyle{statesq}=[shape=rectangle,thick,draw,minimum size=0.6cm]
    \node[statesq,fill={rgb:black,1;white,2}] (c1) {$p_1$};
    \node[statesq,fill={rgb:black,1;white,2}, below left of=c1, yshift=5mm] (b1) {$p_2$};
    \node[state, below of=b2] (Z4) {$c_4$};
    \node[state,below left of=b1] (Z1) {$c_1$};
    \node[state, below of=b1] (Z2) {$c_2$};
    \node[state, below right of=b1] (Z3) {$c_3$};

    \path[->,draw,thick]
    (c1) edge node {} (b1)
    (c1) edge node {} (Z4)
    (b1) edge node {} (Z1)
    (b1) edge node {} (Z2)
    (b1) edge node {} (Z3) ;
  \end{tikzpicture}
   \caption{State 3 $\rightarrow \pmb C_3=f_3(q_2,q_1)$}
\end{subfigure}
 \begin{subfigure}[b]{0.45\textwidth}
   \begin{tikzpicture}[shorten >=1pt,node distance=1.5cm,on grid,auto]
    \tikzstyle{state}=[shape=circle,thick,draw,minimum size=0.6cm]
  \tikzstyle{statesq}=[shape=rectangle,thick,draw,minimum size=0.6cm]
    \node[statesq,fill={rgb:black,1;white,2}] (c1) {$p_1$};
    \node[state, below of=b2] (Z4) {$c_4$};
    \node[state,below left of=b1] (Z1) {$c_1$};
    \node[state, below of=b1] (Z2) {$c_2$};
    \node[state, below right of=b1] (Z3) {$c_3$};

    \path[->,draw,thick]
    (c1) edge node {} (Z4)
    (c1) edge node {} (Z1)
    (c1) edge node {} (Z2)
    (c1) edge node {} (Z3) ;
  \end{tikzpicture}
   \caption{State 2 $\rightarrow \pmb C_2 = f_2(q_1)$}
\end{subfigure}
 \begin{subfigure}[b]{0.45\textwidth}
     \begin{tikzpicture}[shorten >=1pt,node distance=1.5cm,on grid,auto]
    \tikzstyle{state}=[shape=circle,thick,draw,minimum size=0.6cm]
  \tikzstyle{statesq}=[shape=rectangle,thick,draw,minimum size=0.6cm]

    \node[state, below of=b2] (Z4) {$c_4$};
    \node[state,below left of=b1] (Z1) {$c_1$};
    \node[state, below of=b1] (Z2) {$c_2$};
    \node[state, below right of=b1] (Z3) {$c_3$};

    \path[->,draw,thick]
    (Z1)
    (Z2)
    (Z3)
    (Z4) ;
  \end{tikzpicture}
   \caption{State 1 $\rightarrow \pmb C_1 = \pmb I$}
  \end{subfigure}

        \caption{Three parameter graph with three latent variables. Grey and white nodes represent latent factors and observed variables, respectively.}
        \label{fig:pc_corr_ex}
\end{figure}

As shown in Section \ref{sec:sub23} the correlations in a correlation matrix, can be formulated as functions of less parameters based on the graph. To derive the PC prior for the correlation matrix, we adopt and extend the approach used by \cite{sorbye2017penalised} for autoregressive processes. As in the case of an autoregressive process, we can define a sequence of flexible and base models as shown in Section \ref{sec:base} that reduces a given graph of conditional dependence to the independent case in multiple steps. \\ 
To illustrate the derivation of the joint prior, consider again the graphical structure outlined in Figure~\ref{fig:fig3}, as detailed in Section \ref{sec:sub23}. The contraction to the independent case is decomposed into three steps as presented in Figure \ref{fig:pc_corr_ex}. For each step we derive a PC prior for the resulting $\pmb C$. Note that the base model is defined by setting one generating parameter to zero in the flexible model. This implies that we can build up a joint prior by considering multiple univariate PC priors. The joint prior for all the generating parameters of the correlation matrix $\pmb C$ is defined as:

\begin{equation}
    \pi(\pmb C|\lambda) \propto \pi_{4\rightarrow 3}(q_1,q_2,q_3|\lambda) \pi_{3\rightarrow 2}(q_1,q_2|\lambda)\pi_{2\rightarrow 1}(q_1|\lambda).\label{eq:pcpriorC}  
\end{equation}
Following Section \ref{sec:pcprelim} and \cite{sorbye2017penalised}, we can derive the step-specific PC priors based on the KLD between the flexible and base models. The KLD between two Gaussian densities where State 4 denotes the flexible model and State 3 denotes the base model ($q_3 = 0$) is:
\begin{eqnarray*}
    && \text{KLD}(N(\pmb 0, \pmb C_4(q_3))||N(\pmb 0, \pmb C_4(q_3 = 0)) \\
    &=& \text{KLD}(N(\pmb 0, \pmb C_4(q_3))||N(\pmb 0, \pmb C_3)) \\
    &=& \frac{1}{2}\left(tr(\pmb C_3^{-1} \pmb C_4) - n - \log \left(\frac{|\pmb C_4|}{|\pmb C_3|}\right)\right) \\
    &=& f(q_1,q_2,q_3).
\end{eqnarray*}
Based on the KLD, the distance function can be defined as $d(q_1,q_2,q_3) = \sqrt{2f(q_3,q_2,q_1)}$, and  subsequently, the PC prior for the transition from State 4 to State 3 can be derived as
\begin{equation*}
\pi_{4\rightarrow 3}(q_1,q_2,q_3|\lambda) = \lambda \exp\left( -\lambda d(q_1,q_2,q_3)\right)\left|\frac{\partial d(q_1,q_2,q_3)}{\partial q_3}\right|.
\end{equation*}
Similarly, all the PC priors for the different steps are constructed until finally, the joint prior \eqref{eq:pcpriorC} is defined. 
We assume a common rate parameter, $\lambda$, for each step in the sequence. Because the joint prior is built from stepwise model comparisons, we defined one rate parameter that is common to all intermediate steps between $\pmb C$ and $\pmb I$ to ensure equal contraction between steps.\\ 
The sensitivity of the choice of $\lambda$ is investigated in the next section.

\subsection{Choosing the value for \texorpdfstring{$\lambda$}{}} 
\label{sec:prior}

The prior depends on the $\lambda$ parameter, which corresponds to the rate of the exponential prior of the pairwise distance between the base model and its corresponding complex model. 
Close to the base model, the PC prior is a tilted Jeffreys’  where the amount of tilting is determined by the distance on the Riemannian manifold to the base model scaled by the parameter $\lambda$. The user-defined parameter $\lambda$ thus determines the degree of informativeness in the prior. Because the PC priors are defined as weakly informative priors, they tend to be fairly insensitive to the choice of $\lambda$ when there is at least a moderate amount of information. 

We proceed numerically to define a reasonable value for $\lambda$ by sampling distances and mapping them to correlations. The most appropriate value for $\lambda$ is the value that contracts correlation towards 0 but allows for values to be close to 1.

\begin{figure}[h]
\centering
\includegraphics[scale=0.8]{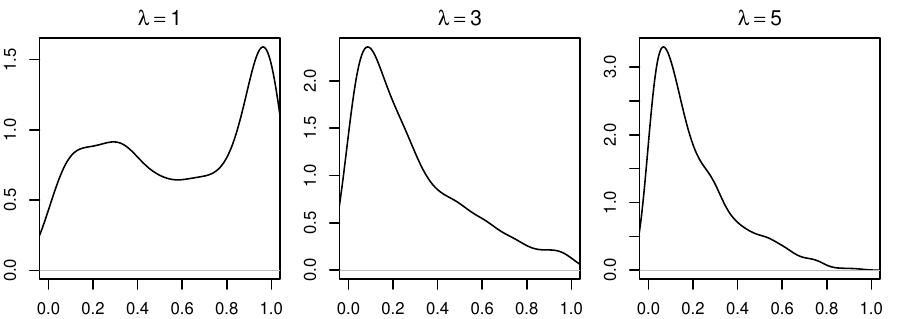}
\caption{Panels show the exponential priors obtained between the flexible model (1 parent) and the base model (0 parents) and three different $\lambda$ values.}
\label{fig:lam1}
\end{figure}

In Figure \ref{fig:lam1}, we illustrate the distribution obtained with different $\lambda$ values for a graph model made of a parent and hence compared with single children, i.e. independent and no parents.
We reach a reasonable contraction around $\lambda=3$, as too small $\lambda$ values do not contract sufficiently towards zero, while too large values have prior probability close to zero for correlations near 1. 

\begin{figure}[h]
\centering
\includegraphics[scale=0.8]{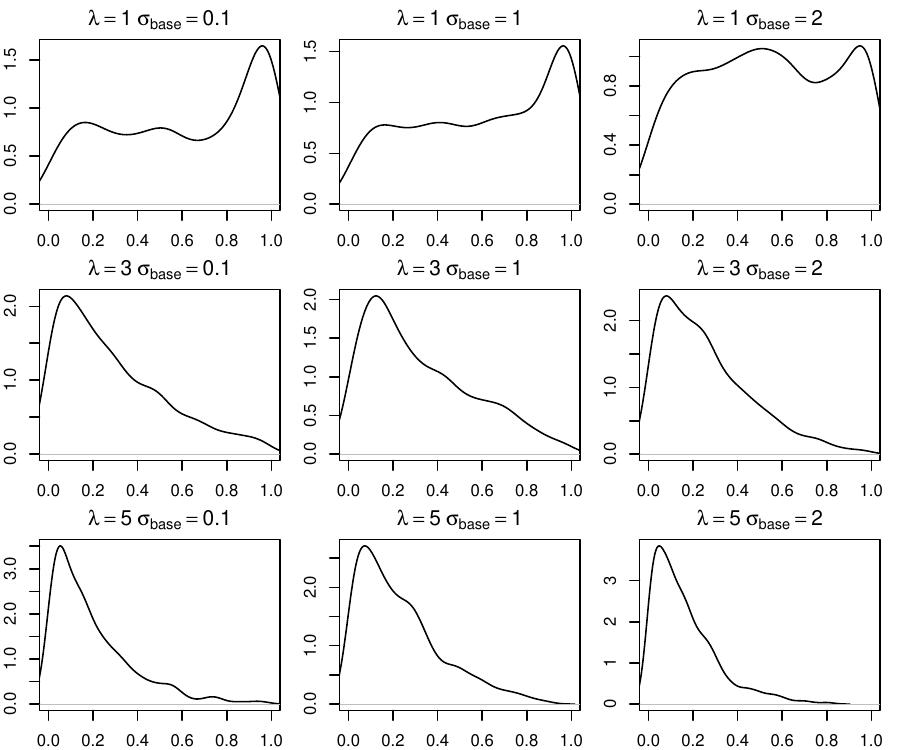}
\caption{Panel shows the exponential priors for the sampling distances between the complex model (2 parents) and the base model (1 parent), computed for $\lambda=(1,3,5)$ and standard deviation of the base model $\sigma_{base}=(0.1,1,2)$.}
\label{fig:lam2}
\end{figure}

In Figure \ref{fig:lam2}, we compare the distribution for $\lambda$ in the case of a complex model made of two parents and a base model of one parent, i.e. exchangeable correlation matrix. In this case, we need to fix the value of the standard deviation of the base model parent. We chose three different values for standard deviation (for interpretability reasons) $\sigma_{base}=(0.1, 1, 2)$ , and $\lambda=3$ provide similar results across the standard deviation values, as well as behaving in a similar way as in Figure \ref{fig:lam1}. 

These examples presented in this section are simple and based on the graph representation, and then we mapped onto the correlation and the different $\lambda$ values. Next, we provide a simulation study on longitudinal joint models, where we explore $\lambda$, misspecified graphs and different sample sizes.

\section{Simulation Studies}
\label{sec:sim}

This section provides two simulation studies for multivariate longitudinal data with 2 and 4 longitudinal markers \cite{verbeke2014analysis, bandyopadhyay2011review}. 
 We define the model as follows: For response variables $\pmb y_{ij}, i = 1,2,\dots,n \quad \text{and} \quad j = 1,2,\dots,p$ define the joint distribution through the univariate conditionals as
\[
 \pi(\pmb y|\pmb \eta, \mm{\theta}) \propto \prod_{i,j} p(y_{ij}|\eta_{ij}, \mm{\theta})
 \]
  where the linear predictors $\pmb\eta$ connects the responses $\pmb y$ to the covariates $\pmb X$ and $\pmb Z$ through fixed effects $\pmb\beta$ and random effects $\pmb b$, respectively, as follows: 
 \begin{equation*}
 \pmb \eta = \pmb X\pmb \beta + \pmb Z\pmb b,
 \end{equation*}
 where we assume Gaussian priors for $\pmb\beta$ and $\pmb b$ with sparse precision matrices. Consider the stacked vector $\mm{x} = (\pmb\beta, \pmb b)$ as the random latent field, then $\pi(\mm{x})$ is a multivariate Gaussian density with a precision matrix composed by the structure of the various effects. All the models were estimated with \textbf{R-INLA}\cite{rue2009approximate,Rue2017,art703}.

\subsection{Multivariate longitudinal model} 

During clinical trials, a group of patients undergoes regular visits, during which data such as blood-measured markers, various tests, and patient-reported outcomes are systematically collected. Our focus lies in modelling several longitudinal markers of interest. Each individual exhibits a distinct deviation from the mean distribution of these markers (random effects), and we posit that there may be potential correlations among the longitudinal markers. 

Typically, a recognised hierarchical structure exists for the correlation between measurements, such as repeated measurements within an individual or a subset of individuals. Consequently, describing the correlation structure of random effects becomes straightforward through a graphical approach.
Let $y_{ijk}$ denote the value of longitudinal marker $k^{th}$ 
for individual $i^{th} (i=1, ..., N_k)$ measured at time points $t_{ijk}$ with $j^{th}$ occasion $(j=1, ..., n_{ik})$. We can describe the distribution of this marker with a mixed effects model:

\[
\textrm{E}(y_{ijk}|\ldots) = \eta_{ik}(t_{ijk}) =\boldsymbol{X}_{ik}(t_{ijk})^\top \boldsymbol{\beta}_k + \boldsymbol{Z}_{ik}(t_{ijk})^\top \boldsymbol{b}_{ik},
\]
where $\eta_{ik}(t_{ijk})$ is the linear predictor defined by fixed effects $\boldsymbol{\beta}_k$ of covariates $\boldsymbol{X}_{ik}(t_{ijk})$ and random effects $\boldsymbol{b}_{ik}$ of covariates $\boldsymbol{Z}_{ik}(t_{ijk})$. The residual error follows a Gaussian distribution, which is assumed negligible in the following. Extension to generalized linear modelling follows trivially.  

\subsubsection{Simulation plan}

In Table \ref{tab:tabsim}, we reported the simulation scenarios, and for each, we simulated 100 datasets and two sample sizes. First, we evaluate the model properties with a low sample size of 30 individuals with 4 repeated measurements each and then a sample size of 1000 individuals with 11 repeated measurements each.

We defined a correlation graph for each scenario and fitted each model with different values of $\lambda=(1,3,5)$ to illustrate the impact on the prior choice. In the first scenario, we included a model with a misspecified hierarchical structure for correlations. Moreover, we added to each scenario an unspecified covariance structure, fitting the model with an Inverse-Wishart prior with matrix $I$ and degrees of freedom 10. and using the mean of the multiple correlation parameters that should match our joint correlations for the comparison of the results. 

\begin{table}[h]
    \centering
\caption{Simulation plan for the longitudinal joint models.}
    \begin{tabular}{ccccc}
    \toprule
Outcomes &  Sample size  & Repeated   measures   & Correlation structure \\
\hline 
Two markers & $N=30 $ & $t=4$  & Figure \ref{figp3c4}, misspecified and unspecified \\ 
    & $N=1000$ &  $t=11$ & \\ 
    \midrule
Four markers  & $N=30$ & $t=4$  & Figure \ref{fig:basemod} and  unspecified \\ 
      & $N=1000$ &  $t=11$\\
\hline
    \end{tabular}
    \label{tab:tabsim}
\end{table}

\subsubsection{Longitudinal joint model: two markers}
\label{sec:timelin}
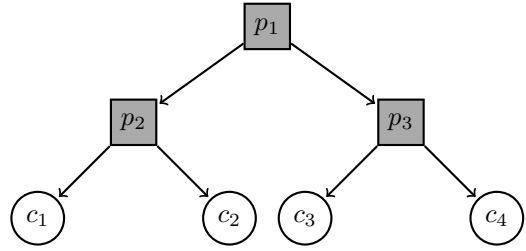
\begin{figure}[h!]
     \centering
        \begin{tikzpicture}[shorten >=1pt,node distance=2.5cm,on grid,auto]
    \tikzstyle{state}=[shape=circle,thick,draw,minimum size=0.6cm]
    \tikzstyle{statesq}=[shape=rectangle,thick,draw,minimum size=0.6cm]
    \node[statesq,fill={rgb:black,1;white,2}] (p1) {$p_1$};
    \node[statesq,fill={rgb:black,1;white,2}, below left of=p1, yshift=5mm] (p2) {$p_2$};
    \node[statesq,fill={rgb:black,1;white,2}, below right of=p1, yshift=5mm] (p3) {$p_3$};
    \node[state, below left of=p2, xshift=5mm, yshift=5mm] (c1) {$c_1$};
    \node[state, below right of=p2, xshift=-5mm, yshift=5mm] (c2) {$c_2$};
    \node[state,below left of=p3, xshift=5mm, yshift=5mm] (c3) {$c_3$};
    \node[state, below right of=p3, xshift=-5mm, yshift=5mm] (c4) {$c_4$};
    \path[->,draw,thick]
    (p1) edge node {} (p2)
    (p1) edge node {} (p3)
    (p2) edge node {} (c1)
    (p2) edge node {} (c2)
    (p3) edge node {} (c3)
    (p3) edge node {} (c4);
  \end{tikzpicture}
        \caption{Graph for the model fitted on longitudinal data in Section \ref{sec:timelin} with 3 parents and 4 children. Children 1 and 2 correspond to random intercept and slope for the first marker, while children 3 and 4 correspond to random intercept and slope for the second marker.}
         \label{figp3c4}
\end{figure}

We define the two markers model as Gaussian data and a linear temporal effect:
\begin{equation*}
\begin{split}
    \textrm{E}(y_{ijk} | \ldots) &= \beta_{0k} + b_{0ik} + (\beta_{1k} + b_{1ik})t_{ijk}.
\end{split}
\end{equation*}
Here $\beta_{0k}$ and $\beta_{1k}$ are the fixed intercept and slope for marker $k$, respectively, while $b_{0ik}$ and $b_{1ik}$ are the corresponding random intercept and slope. The model includes 4 random effects (random intercept and slope for each marker). We assume the structure described in Figure \ref{figp3c4}.

In this graph, $c_1$ and $c_2$ represent the random intercept and slope of the first longitudinal marker, and $c_3$ and $c_4$ represent the random intercept and slope of the second marker. One can deduce a hierarchy in the correlations between these random effects as $c_1$ and $c_2$ share the common ancestors $p_2$ and $p_1$, and similarly, the other effects shares $p_1$ and $p_3$. The correlation between the two subgroups of children is quantified by their common ancestor $p_1$ and is lower than the correlations between random intercepts and slopes. The correlation structure is then defined as follows:\\
\begin{equation*}
\left.
\begin{NiceArray}{ll:ll}
\sigma_{c1} & \\
\rho_1 & \sigma_{c2} \\
\hdashline
\multicolumn{2}{c}{\multirow{2}{*}{$\rho_3$}} & \sigma_{c3}\\
 & & \rho_2 & \sigma_{c4}
\end{NiceArray} \right.
\end{equation*}
The value of the fixed intercept and slope for each longitudinal marker is defined as $\beta_{0k}=0.2$ and $\beta_{1k}=-0.1$, respectively, while we fixed the correlation between the subgroups of children to 0.9 and the correlation between the two longitudinal markers to 0.8. 

Based on the graph provided, we notice that the model balances an unspecified covariance structure and a fully independent structure by defining single correlation parameters between groups of random effects. In the given example, the correlation parameter $\rho_3$ measures the correlation between two sets of correlated random effects, corresponding to the random intercept and slope for each longitudinal marker. An unspecified correlation would necessitate four correlation parameters instead of one, accounting for the relationship between the two longitudinal markers. This structure facilitates a more intuitive interpretation of the correlation between markers, as it is quantified by a single parameter. In contrast, an unspecified structure involves four terms that could have opposite signs, making the interpretation of the correlation between markers challenging.

Furthermore, to investigate the results of an ill-defined model, we fitted a misspecified model where we swapped $c_2$ and $c_4$ so that the hierarchy of correlations is not correct (i.e., we assume the correlation between $c_1$ and $c_2$ is lower than between $c_1$ and $c_4$ while it is higher).

\begin{table}[!ht]
\caption{Simulations results with 3 parents and 4 children for $\lambda=1, 3$ and $5$. The column ``misspecified'' corresponds to a model with the wrong hierarchical order for the correlation structure of children (using $\lambda=3$), and the column ``unspecified'' corresponds to the unspecified correlation between all children (results presented as mean [95\% CI]). Computation time is provided in seconds in the last row.}
\footnotesize
{\tabcolsep=2pt \renewcommand{\arraystretch}{1.5}
\begin{tabular}{@{}lccccccccccccccccccccccccccccc@{}}
\hline
True values & $\lambda$ = 1 & $\lambda$ = 3 & $\lambda$ = 5 & misspecified & unspecified\\
\hline
\multicolumn{6}{c}{N=30 individuals}\\\hline
$\sigma_{c1}$=1 & 1.03 [0.79, 1.3] & 0.99 [0.76, 1.25] & 0.95 [0.74, 1.2] & 1 [0.77, 1.26] & 0.98 [0.77, 1.23] & \\
 $\sigma_{c2}$=0.2 & 0.21 [0.16, 0.26] & 0.2 [0.15, 0.24] & 0.19 [0.15, 0.24] & 0.2 [0.15, 0.25] & 0.27 [0.23, 0.3] & \\
 $\sigma_{c3}$=0.1 & 0.1 [0.08, 0.12] & 0.1 [0.08, 0.12] & 0.09 [0.07, 0.12] & 0.1 [0.08, 0.12] & 0.21 [0.2, 0.22] & \\
 $\sigma_{c4}$=0.5 & 0.51 [0.39, 0.63] & 0.49 [0.38, 0.6] & 0.47 [0.36, 0.58] & 0.5 [0.39, 0.61] & 0.51 [0.41, 0.62] & \\
 $\rho_1$=0.9 & 0.9 [0.84, 0.95] & 0.89 [0.81, 0.94] & 0.87 [0.78, 0.93] & 0.81 [0.69, 0.89] & 0.65 [0.51, 0.76] & \\
 $\rho_2$=0.9 & 0.91 [0.85, 0.95] & 0.89 [0.83, 0.94] & 0.87 [0.8, 0.93] & 0.81 [0.69, 0.89] & 0.4 [0.29, 0.49] & \\
 $\rho_3$=0.8 & 0.8 [0.66, 0.89] & 0.79 [0.64, 0.89] & 0.78 [0.62, 0.88] & 0.81 [0.69, 0.89] & 0.48 [0.37, 0.59] & \\
 \midrule $\beta_{01}$=0.2 & 0.19 [-0.19, 0.49] & 0.19 [-0.19, 0.49] & 0.19 [-0.19, 0.49] & 0.19 [-0.19, 0.49] & 0.19 [-0.19, 0.49] & \\
 $\beta_{11}$=-0.1 & -0.1 [-0.16, -0.04] & -0.1 [-0.16, -0.04] & -0.1 [-0.16, -0.04] & -0.1 [-0.16, -0.04] & -0.1 [-0.16, -0.04] & \\
 $\beta_{02}$=0.2 & 0.2 [0.16, 0.23] & 0.2 [0.16, 0.23] & 0.2 [0.16, 0.23] & 0.2 [0.16, 0.23] & 0.2 [0.17, 0.23] & \\
 $\beta_{12}$=-0.1 & -0.1 [-0.28, 0.07] & -0.1 [-0.28, 0.07] & -0.1 [-0.28, 0.07] & -0.1 [-0.28, 0.07] & -0.1 [-0.28, 0.07] & \\
 \midrule 
 Comp. time (s) & 1.11 [0.87, 1.48] & 0.94 [0.86, 1.17] & 0.98 [0.85, 1.17] & 0.96 [0.88, 1.22] & 1.88 [1.61, 2.3] & \\
\hline
\multicolumn{6}{c}{N=1000 individuals}\\ \hline
$\sigma_{c1}$=1 & 1 [0.95, 1.04] & 1 [0.94, 1.04] & 0.99 [0.94, 1.04] & 1 [0.94, 1.04] & 1 [0.95, 1.04] & \\
 $\sigma_{c2}$=0.2 & 0.2 [0.19, 0.21] & 0.2 [0.19, 0.21] & 0.2 [0.19, 0.21] & 0.2 [0.19, 0.21] & 0.2 [0.19, 0.21] & \\
 $\sigma_{c3}$=0.1 & 0.1 [0.1, 0.1] & 0.1 [0.1, 0.1] & 0.1 [0.1, 0.1] & 0.1 [0.1, 0.1] & 0.1 [0.1, 0.11] & \\
 $\sigma_{c4}$=0.5 & 0.5 [0.48, 0.52] & 0.5 [0.47, 0.52] & 0.5 [0.47, 0.52] & 0.5 [0.48, 0.52] & 0.5 [0.48, 0.52] & \\
 $\rho_1$=0.9 & 0.9 [0.89, 0.91] & 0.9 [0.89, 0.91] & 0.9 [0.89, 0.91] & 0.83 [0.82, 0.85] & 0.89 [0.88, 0.9] & \\
 $\rho_2$=0.9 & 0.9 [0.89, 0.91] & 0.9 [0.89, 0.91] & 0.9 [0.89, 0.91] & 0.83 [0.82, 0.85] & 0.85 [0.84, 0.87] & \\
 $\rho_3$=0.8 & 0.8 [0.78, 0.82] & 0.8 [0.78, 0.82] & 0.8 [0.78, 0.82] & 0.83 [0.82, 0.85] & 0.77 [0.76, 0.79] & \\
 \midrule $\beta_{01}$=0.2 & 0.2 [0.14, 0.26] & 0.2 [0.14, 0.26] & 0.2 [0.14, 0.26] & 0.2 [0.14, 0.26] & 0.2 [0.14, 0.26] & \\
 $\beta_{11}$=-0.1 & -0.1 [-0.11, -0.09] & -0.1 [-0.11, -0.09] & -0.1 [-0.11, -0.09] & -0.1 [-0.11, -0.09] & -0.1 [-0.11, -0.09] & \\
 $\beta_{02}$=0.2 & 0.2 [0.2, 0.21] & 0.2 [0.2, 0.21] & 0.2 [0.2, 0.21] & 0.2 [0.2, 0.21] & 0.2 [0.2, 0.21] & \\
 $\beta_{12}$=-0.1 & -0.1 [-0.13, -0.07] & -0.1 [-0.13, -0.07] & -0.1 [-0.13, -0.07] & -0.1 [-0.13, -0.07] & -0.1 [-0.13, -0.07] & \\
 \midrule 
 Comp. time (s) & 20.68 [17.12, 24.74] & 20.41 [16.37, 25.47] & 20.63 [17.72, 25.38] & 24.51 [18.21, 38.67] & 25.46 [21.54, 29.88]]\\
\hline
\end{tabular}}
\label{Ex3p4c}
\end{table}

The results of the simulations are presented in Table \ref{Ex3p4c}, estimates and 95\% credible intervals (CI). Our results show the good performances of the model to capture true values of parameters used to simulate data with the chosen prior $\lambda=3$. Negligible differences are observed when using a lower value of $\lambda$ as the contraction towards zero property is not influential when posteriors are driven by the data. Using a higher value of $\lambda$ slightly shifts parameter estimates down when sample size is low as this prior is more informative and conflicts with data information (i.e., low prior probability of high correlation while the true correlations are high) but does not change the posteriors when more data information is available.

When the model is misspecified, i.e., where we swapped $c_2$ and $c_4$ (from Figure \ref{figp3c4}), the correlation between $c_1$ and $c_4$ is constrained to be higher than the correlation between $c_1$ and $c_2$ while it is supposed to be lower. This model gives the same correlation for all $\rho_1$, $\rho_2$ and $\rho_3$, indicating the model may not be properly specified. Indeed, when we observe the exact same correlation from different parameters (values presented in tables are rounded but the posteriors remain identical with more digits), it could be because these correlations are similar (in this case there are some useless parameters) or because the assumed graph structure is not correct (i.e., wrong hierarchy). The model with unspecified covariance structure was not able to recover the true correlations with low sample size but found more accurate results with the larger sample size, although not as accurate as our new approach. Since this model is fairly simple and there are only 3 less hyperparameters with our method, the gain in computation time is minor as all models fitted within a few seconds.

\subsubsection{Longitudinal joint model: four markers}
\label{sec:longiEx2}

To evaluate how the methods proposed scales for more complex problems. We define a model with eight random effects (intercepts and slopes), represented in graph $a$ from Figure \ref{fig:basemod}.  
We also include two fixed effects associated with a binary ($X_{bin} \sim Binomial(0.5)$) and a continuous ($X_{con} \sim \mathcal{N}(0,0.5)$) covariate, as follows: 
\[ 
y_{ijk} = \beta_{0k} + b_{0ik} + (\beta_{1k} + b_{1ik})t_{ijk} + 
\beta_{2k} X_{bin} + \beta_{3k} X_{con} + \sigma_{\varepsilon k}, \\
 k=1,2,3,4.
 \]
Following the derivation described in Section\ref{sec:comp}, the parents define the correlation between each random effect, giving the following correlation structure:
\begin{equation*}
\left.
\begin{NiceArray}{cc:cc:cc:cc}
\sigma_{c1} & \\
\rho_1 & \sigma_{c2} \\
\hdashline
\multicolumn{2}{c}{\multirow{2}{*}{$\rho_5$}} & \sigma_{c3}\\
 & & \rho_2 & \sigma_{c4}\\
 \hdashline
\multicolumn{2}{c}{\multirow{2}{*}{$\rho_7$}} & \multicolumn{2}{c}{\multirow{2}{*}{$\rho_9$}} & \sigma_{c5}\\
 & & & & \rho_3 & \sigma_{c6}\\
 \hdashline
\multicolumn{2}{c}{\multirow{2}{*}{$\rho_8$}} & \multicolumn{2}{c}{\multirow{2}{*}{$\rho_{10}$}} & \multicolumn{2}{c}{\multirow{2}{*}{$\rho_6$}} & \sigma_{c7}\\
& & & & & & \rho_4 & \sigma_{c8}\\
\end{NiceArray} \right.
\end{equation*}

As for the previous example, in this graph, $c_1$ and $c_2$ represent the random intercept and slope of the first longitudinal marker, and $c_3$ and $c_4$ represent the random intercept and slope of the second marker, and so on.  One can deduce a hierarchy in the correlations between these random effects as $c_1$ and $c_2$ share the common ancestors $p_1$, $p_2$, and $p_4$ and similarly, the other effect shares $p_1$,$p_2$,  and $p_5$ for $c_3$ and $c_4$, etc. 

The results are reported in Tables \ref{Ex7p8c} and \ref{Ex7p8c2} for $N=30$ and $N=1000$, respectively. 
In this more complex scenario,  the choice of $\lambda$ can slightly impact results when there is little data information, while it has a negligible impact for a larger sample size. 
In fact, for $N=30$, $\lambda=1$ leads to a contraction of correlation priors towards one and tends to overestimate covariances, while a  $\lambda=5$ value has a strong contraction towards 0, which leads to underestimation of covariances when data information is not enough. 
With more data available, ($N=1000$) the choice of the $\lambda$ has no effect as the contraction towards zero property is not influential when posteriors are driven by the data, as seen for the two-makers simulations.  Overall,  PC priors are fairly insensitive to the choice of $\lambda$ when there is at least a moderate amount of information \cite{Simpson2014}. In line with the results from the first scenario, the model with an unspecified structure is clearly outperformed by our approach for low sample size while it performs similar when data is sufficiently informative.

Finally, we included the computation times in each table. With $N=30$, our proposed approach is nearly 10 times faster than the unspecified structure while for the simulations with $N=1000$, computation times were nearly 6 times faster. Our approach indeed requires the estimation of 18 parameters for the covariance structure of the random effects (8 random-effects specific standard deviation terms and 10 correlation terms) while the unstructured approach requires 36 parameters (8  standard deviations and 28 correlation).

\subsubsection{Discussion of the simulations results}
\label{sec:discsim}

Our simulations illustrate some key features of the proposed approach. With a low sample size, priors have an impact on posteriors and   our proposal outperforms the traditional unstructured covariance approach in terms of accuracy and stability of the correlation estimates. 
The choice of the $\lambda$ can slightly influence the results when the sample size is low and prior information conflicts with data information (i.e., $\lambda$ too low/high will shift posterior correlations towards the edges). With a larger sample size, under the assumption that the graph is correct,  priors have a minor impact on the results, and the main difference is in computation time, where our proposed approach is much faster due to the reduced number of parameters to be estimated. 

 In the first scenario, we included the misspecified case, where we have assumed hierarchically structured correlations. However, the estimated correlations were all the same. This information can warn the user about possible issues in the model structure. Hence, it could guide the order of the graph through a stepwise procedure. For example, if one does not know whether correlation $\rho_1$ is expected to be higher/lower compared to correlation $\rho_2$, running two models will allow identifying the correct structure as one of the models should ``hit a wall'' and return identical correlations for $\rho_1$ and $\rho_2$. 
 
 Further, we re-run all the simulation scenarios for lower correlation values ($<0.5$), see Tables \ref{tab:APPEx3p4c},\ref{tab:APPEx7p8c}, \ref{tab:APPEx7p8c2},  in Appendix \ref{sec:appC} We have drawn the same conclusions as for the higher correlation presented in this section.

\begin{table}[ht]
\caption{Simulations results with 7 parents and 8 children for $\lambda=1, 3$ and $5$ with 30 individuals per dataset. The column ``unspecified'' corresponds to the unspecified correlation between all children (results presented as: mean [95\% CI]). Computation time is provided in seconds in the last row.}
\footnotesize
\centering
{\tabcolsep=2pt \renewcommand{\arraystretch}{1.5}
\begin{tabular}{@{}lccccccccccccccccccccccccccccc@{}}
\hline
\multicolumn{5}{c}{N=30 individuals}\\
True values & $\lambda$ = 1 & $\lambda$ = 3 & $\lambda$ = 5 & unspecified \\
\hline
$\sigma_{c1}$=1 & 1.02 [0.77, 1.26] & 0.95 [0.72, 1.17] & 0.9 [0.68, 1.11] & 0.95 [0.71, 1.15] & \\
 $\sigma_{c2}$=0.5 & 0.51 [0.37, 0.66] & 0.48 [0.35, 0.62] & 0.45 [0.33, 0.58] & 0.5 [0.38, 0.64] & \\
 $\sigma_{c3}$=1 & 1.04 [0.77, 1.28] & 0.96 [0.72, 1.19] & 0.91 [0.68, 1.12] & 0.96 [0.7, 1.2] & \\
 $\sigma_{c4}$=0.5 & 0.52 [0.4, 0.64] & 0.48 [0.37, 0.59] & 0.46 [0.36, 0.56] & 0.51 [0.41, 0.62] & \\
 $\sigma_{c5}$=1 & 1.08 [0.81, 1.36] & 1 [0.76, 1.26] & 0.95 [0.72, 1.19] & 0.98 [0.74, 1.23] & \\
 $\sigma_{c6}$=0.5 & 0.54 [0.41, 0.69] & 0.5 [0.38, 0.64] & 0.47 [0.36, 0.61] & 0.51 [0.41, 0.64] & \\
 $\sigma_{c7}$=1 & 1.05 [0.81, 1.34] & 0.98 [0.75, 1.25] & 0.92 [0.71, 1.17] & 0.96 [0.74, 1.22] & \\
 $\sigma_{c8}$=0.5 & 0.52 [0.38, 0.68] & 0.49 [0.36, 0.63] & 0.46 [0.34, 0.59] & 0.5 [0.4, 0.63] & \\
 $\rho_1$=0.9 & 0.9 [0.83, 0.95] & 0.87 [0.79, 0.93] & 0.84 [0.75, 0.91] & 0.82 [0.69, 0.9] & \\
 $\rho_2$=0.9 & 0.91 [0.84, 0.95] & 0.88 [0.8, 0.93] & 0.85 [0.76, 0.91] & 0.82 [0.7, 0.9] & \\
 $\rho_3$=0.9 & 0.91 [0.86, 0.95] & 0.89 [0.82, 0.94] & 0.86 [0.77, 0.92] & 0.83 [0.73, 0.9] & \\
 $\rho_4$=0.9 & 0.91 [0.86, 0.95] & 0.89 [0.82, 0.94] & 0.86 [0.77, 0.92] & 0.82 [0.73, 0.9] & \\
 $\rho_5$=0.8 & 0.8 [0.68, 0.88] & 0.78 [0.65, 0.87] & 0.76 [0.61, 0.85] & 0.74 [0.59, 0.85] & \\
 $\rho_6$=0.8 & 0.82 [0.71, 0.9] & 0.79 [0.67, 0.89] & 0.77 [0.63, 0.87] & 0.74 [0.59, 0.85] & \\
 $\rho_7$=0.6 & 0.64 [0.43, 0.79] & 0.61 [0.4, 0.77] & 0.58 [0.36, 0.74] & 0.55 [0.34, 0.71] & \\
 $\rho_8$=0.6 & 0.63 [0.45, 0.78] & 0.6 [0.41, 0.76] & 0.58 [0.38, 0.74] & 0.55 [0.35, 0.73] & \\
 $\rho_9$=0.6 & 0.64 [0.42, 0.77] & 0.61 [0.38, 0.75] & 0.58 [0.35, 0.73] & 0.55 [0.31, 0.71] & \\
 $\rho_{10}$=0.6 & 0.63 [0.42, 0.77] & 0.6 [0.38, 0.75] & 0.57 [0.35, 0.73] & 0.55 [0.3, 0.72] & \\
 \midrule $\beta_{01}$=0.2 & 0.15 [-0.29, 0.64] & 0.15 [-0.29, 0.66] & 0.15 [-0.29, 0.66] & 0.15 [-0.35, 0.7] & \\
 $\beta_{11}$=-0.1 & -0.11 [-0.27, 0.09] & -0.11 [-0.27, 0.09] & -0.11 [-0.27, 0.09] & -0.11 [-0.27, 0.09] & \\
 $\beta_{21}$=-0.2 & -0.2 [-0.52, 0.07] & -0.2 [-0.52, 0.07] & -0.2 [-0.52, 0.07] & -0.2 [-0.53, 0.07] & \\
 $\beta_{31}$=0.1 & 0.13 [-0.16, 0.41] & 0.13 [-0.15, 0.42] & 0.13 [-0.17, 0.42] & 0.13 [-0.2, 0.46] & \\
 $\beta_{02}$=0.2 & 0.17 [-0.38, 0.67] & 0.17 [-0.38, 0.66] & 0.17 [-0.38, 0.66] & 0.17 [-0.38, 0.66] & \\
 $\beta_{12}$=-0.1 & -0.11 [-0.3, 0.04] & -0.11 [-0.3, 0.04] & -0.11 [-0.3, 0.04] & -0.11 [-0.3, 0.04] & \\
 $\beta_{22}$=-0.2 & -0.2 [-0.49, 0.11] & -0.19 [-0.5, 0.11] & -0.19 [-0.51, 0.11] & -0.18 [-0.53, 0.14] & \\
 $\beta_{32}$=0.1 & 0.12 [-0.17, 0.47] & 0.12 [-0.17, 0.47] & 0.11 [-0.19, 0.48] & 0.11 [-0.18, 0.47] & \\
 $\beta_{03}$=0.2 & 0.19 [-0.33, 0.71] & 0.19 [-0.34, 0.71] & 0.19 [-0.36, 0.7] & 0.19 [-0.4, 0.74] & \\
 $\beta_{13}$=-0.1 & -0.1 [-0.26, 0.06] & -0.1 [-0.26, 0.06] & -0.1 [-0.26, 0.06] & -0.1 [-0.26, 0.06] & \\
 $\beta_{23}$=-0.2 & -0.21 [-0.45, 0.1] & -0.21 [-0.46, 0.11] & -0.21 [-0.46, 0.12] & -0.21 [-0.46, 0.14] & \\
 $\beta_{33}$=0.1 & 0.11 [-0.19, 0.48] & 0.11 [-0.21, 0.48] & 0.11 [-0.22, 0.46] & 0.11 [-0.24, 0.45] & \\
 $\beta_{04}$=0.2 & 0.2 [-0.32, 0.74] & 0.2 [-0.3, 0.75] & 0.2 [-0.3, 0.76] & 0.2 [-0.3, 0.78] & \\
 $\beta_{14}$=-0.1 & -0.11 [-0.27, 0.11] & -0.11 [-0.27, 0.11] & -0.11 [-0.27, 0.11] & -0.11 [-0.27, 0.11] & \\
 $\beta_{24}$=-0.2 & -0.21 [-0.52, 0.08] & -0.21 [-0.51, 0.08] & -0.21 [-0.5, 0.07] & -0.22 [-0.53, 0.07] & \\
 $\beta_{34}$=0.1 & 0.1 [-0.22, 0.45] & 0.1 [-0.21, 0.44] & 0.1 [-0.21, 0.44] & 0.11 [-0.24, 0.46] & \\
 \midrule
 Comp. time (s) & 2.68 [2.27, 3.17] & 2.75 [2.37, 3.22] & 2.71 [2.37, 3.29] & 25.66 [21.68, 30.83] & \\
 \hline
\end{tabular}}
\label{Ex7p8c}
\end{table}
\clearpage

\begin{table}[!ht]
\caption{Simulations results with 7 parents and 8 children for $\lambda=1, 3$ and 5 with 1000 individuals per dataset. The column ``unspecified'' corresponds to the unspecified correlation between all children (results presented as mean [95\% CI]). Computation time is provided in seconds in the last row.}
\footnotesize
\centering
{\tabcolsep=2pt \renewcommand{\arraystretch}{1.5}
\begin{tabular}{@{}lccccccccccccccccccccccccccccc@{}}
\hline
 \multicolumn{5}{c}{N=1000 individuals}\\
True values & $\lambda$ = 1 & $\lambda$ = 3 & $\lambda$ = 5 & unspecified \\
\hline
$\sigma_{c1}$=1 & 1 [0.96, 1.04] & 1 [0.96, 1.04] & 0.99 [0.96, 1.04] & 1 [0.97, 1.04] & \\
 $\sigma_{c2}$=0.5 & 0.5 [0.48, 0.52] & 0.5 [0.48, 0.52] & 0.5 [0.47, 0.52] & 0.5 [0.48, 0.52] & \\
 $\sigma_{c3}$=1 & 1 [0.95, 1.04] & 0.99 [0.94, 1.04] & 0.99 [0.94, 1.03] & 1 [0.96, 1.04] & \\
 $\sigma_{c4}$=0.5 & 0.5 [0.47, 0.52] & 0.5 [0.47, 0.52] & 0.5 [0.47, 0.52] & 0.5 [0.48, 0.52] & \\
 $\sigma_{c5}$=1 & 1 [0.95, 1.04] & 0.99 [0.95, 1.04] & 0.99 [0.94, 1.04] & 1 [0.94, 1.05] & \\
 $\sigma_{c6}$=0.5 & 0.5 [0.48, 0.52] & 0.5 [0.48, 0.52] & 0.5 [0.48, 0.52] & 0.5 [0.47, 0.53] & \\
 $\sigma_{c7}$=1 & 1 [0.96, 1.05] & 0.99 [0.95, 1.05] & 0.99 [0.95, 1.05] & 0.99 [0.94, 1.04] & \\
 $\sigma_{c8}$=0.5 & 0.5 [0.47, 0.52] & 0.5 [0.47, 0.52] & 0.5 [0.47, 0.52] & 0.5 [0.47, 0.52] & \\
 $\rho_1$=0.9 & 0.9 [0.89, 0.91] & 0.9 [0.89, 0.91] & 0.9 [0.89, 0.91] & 0.9 [0.88, 0.91] & \\
 $\rho_2$=0.9 & 0.9 [0.89, 0.91] & 0.9 [0.89, 0.91] & 0.9 [0.89, 0.91] & 0.9 [0.88, 0.91] & \\
 $\rho_3$=0.9 & 0.9 [0.89, 0.91] & 0.9 [0.89, 0.91] & 0.9 [0.89, 0.91] & 0.9 [0.89, 0.91] & \\
 $\rho_4$=0.9 & 0.9 [0.89, 0.91] & 0.9 [0.89, 0.91] & 0.9 [0.89, 0.91] & 0.9 [0.89, 0.91] & \\
 $\rho_5$=0.8 & 0.8 [0.78, 0.82] & 0.8 [0.78, 0.82] & 0.8 [0.78, 0.82] & 0.8 [0.78, 0.81] & \\
 $\rho_6$=0.8 & 0.8 [0.78, 0.82] & 0.8 [0.78, 0.82] & 0.8 [0.78, 0.82] & 0.8 [0.78, 0.82] & \\
 $\rho_7$=0.6 & 0.6 [0.56, 0.64] & 0.6 [0.56, 0.64] & 0.6 [0.56, 0.64] & 0.59 [0.57, 0.63] & \\
 $\rho_8$=0.6 & 0.6 [0.56, 0.64] & 0.6 [0.56, 0.63] & 0.6 [0.56, 0.63] & 0.59 [0.57, 0.63] & \\
 $\rho_9$=0.6 & 0.6 [0.56, 0.64] & 0.6 [0.56, 0.64] & 0.6 [0.56, 0.64] & 0.59 [0.56, 0.63] & \\
 $\rho_{10}$=0.6 & 0.6 [0.57, 0.64] & 0.6 [0.57, 0.64] & 0.6 [0.57, 0.64] & 0.59 [0.56, 0.63] & \\
 \midrule $\beta_{01}$=0.2 & 0.2 [0.11, 0.26] & 0.2 [0.11, 0.26] & 0.2 [0.11, 0.26] & 0.2 [0.14, 0.24] & \\
 $\beta_{11}$=-0.1 & -0.1 [-0.13, -0.07] & -0.1 [-0.13, -0.07] & -0.1 [-0.13, -0.07] & -0.1 [-0.12, -0.08] & \\
 $\beta_{21}$=-0.2 & -0.2 [-0.25, -0.14] & -0.2 [-0.25, -0.14] & -0.2 [-0.25, -0.14] & -0.19 [-0.23, -0.14] & \\
 $\beta_{31}$=0.1 & 0.1 [0.05, 0.14] & 0.1 [0.05, 0.14] & 0.1 [0.05, 0.14] & 0.09 [0.06, 0.13] & \\
 $\beta_{02}$=0.2 & 0.2 [0.11, 0.28] & 0.2 [0.11, 0.28] & 0.2 [0.11, 0.28] & 0.21 [0.14, 0.27] & \\
 $\beta_{12}$=-0.1 & -0.1 [-0.13, -0.07] & -0.1 [-0.13, -0.07] & -0.1 [-0.13, -0.07] & -0.1 [-0.12, -0.07] & \\
 $\beta_{22}$=-0.2 & -0.2 [-0.24, -0.15] & -0.2 [-0.24, -0.15] & -0.2 [-0.24, -0.15] & -0.19 [-0.23, -0.15] & \\
 $\beta_{32}$=0.1 & 0.1 [0.05, 0.15] & 0.1 [0.05, 0.15] & 0.1 [0.05, 0.15] & 0.09 [0.05, 0.13] & \\
 $\beta_{03}$=0.2 & 0.2 [0.12, 0.28] & 0.2 [0.12, 0.28] & 0.2 [0.12, 0.28] & 0.19 [0.1, 0.28] & \\
 $\beta_{13}$=-0.1 & -0.1 [-0.13, -0.07] & -0.1 [-0.13, -0.07] & -0.1 [-0.13, -0.07] & -0.1 [-0.12, -0.08] & \\
 $\beta_{23}$=-0.2 & -0.2 [-0.25, -0.15] & -0.2 [-0.25, -0.15] & -0.2 [-0.25, -0.15] & -0.21 [-0.25, -0.18] & \\
 $\beta_{33}$=0.1 & 0.1 [0.05, 0.15] & 0.1 [0.05, 0.15] & 0.1 [0.05, 0.15] & 0.11 [0.05, 0.16] & \\
 $\beta_{04}$=0.2 & 0.2 [0.12, 0.28] & 0.2 [0.12, 0.28] & 0.2 [0.12, 0.28] & 0.18 [0.1, 0.24] & \\
 $\beta_{14}$=-0.1 & -0.1 [-0.12, -0.07] & -0.1 [-0.12, -0.07] & -0.1 [-0.12, -0.07] & -0.1 [-0.12, -0.07] & \\
 $\beta_{24}$=-0.2 & -0.2 [-0.26, -0.14] & -0.2 [-0.26, -0.14] & -0.2 [-0.26, -0.14] & -0.2 [-0.24, -0.16] & \\
 $\beta_{34}$=0.1 & 0.1 [0.05, 0.15] & 0.1 [0.05, 0.15] & 0.1 [0.05, 0.15] & 0.11 [0.07, 0.16] & \\
 \midrule 
 Comp. time (s) & 75.12 [63.95, 88.3] & 77.05 [62.25, 89.23] & 75.81 [65.12, 89.76] & 484.94 [441.66, 504.18] & \\
 \hline
\end{tabular}}
\label{Ex7p8c2}
\end{table}
\clearpage

\section{Application to real data}
\label{sec:data}

We provide two data examples:(i) a multivariate model for four epigenetic biomarkers, where we are able to recover the observed correlation and to define a graph based on the epigenetic potential features. (ii) A multivariate disease mapping on four types of cancer counts in Germany We fit five models with different combinations of disease and spatial correlation: independent,  unstructured correlation, Besag York Mollie \cite{Besag1974} and user-specified graphs. 

\subsection{Simple multivariate model}
\label{sec:modsec}

We illustrate the simple case of multivariate penalized prior by analysing a subset of data from the Irish Longitudinal Study on Ageing (TILDA) cohort \cite{mccrory2021grimage}. The TILDA cohort investigates the physical, mental health, and cognitive measures of Irish residents over 50 years old \cite{kearney2011cohort,whelan2013design}. 
The multivariate outcome is given by four epigenetic biomarkers: Horvath, Hannum, PhenoAge and GrimAge. The epigenetic clock measures an individual's chronological and biological age. Individuals with good lifestyles (no smoking, limited drinking, active lifestyle), are usually genetically younger than their chronological age, and vice versa. Horvath and Hannum are described as the ``first generation clock'' as both have been trained on blood samples and aimed at clocks to predict human chronological age. In contrast, PhenoAge and GrimAge are known as the ``second generation clock'', as both were trained using other biomarkers and mortality data to reflect ageing-related physiological conditions.

Based on this prior knowledge, we can design the correlation structure of these 4 outcomes using a graph as described in Figure \ref{figp3c4}. The first-generation clock Horvath (children 1) and Hannum (children 2) share some variability captured by their common ancestors' parent 2 ($p_2$) and parent 1 ($p_1$), while the second-generation clock shares some variability captured by their common ancestors' parent 3 ($p_3$) and parent 1. The correlation between first and second-generation clocks is constrained to be lower because they only share parent 1 as a common ancestor. This structure reduces the number of parameters to be estimated from 6 correlations to only 3 compared to an unspecified covariance structure, as there is only one correlation parameter between the two groups of children. 
As described in Section \ref{sec:base}, the prior is based on the pairwise sequential distance from the complex model described in Figure \ref{figp3c4}, until reaching the base model where all children are independent. The observed biomarkers show a correlation of $0.85$ and $0.49$ for first and second generation, respectively (see Table \ref{tab:tab1}). 
We fit a simple model with no covariates to compare with the correlations provided by our model:
\[
y_{ij}=\alpha+x_{ij}, \quad i=1,...,215, j=1,...,4
\]

\begin{table}[h]
\centering
\caption{Empirical epigenetic biomarkers correlation.}
\label{tab:tab1}
\begin{tabular}{rrrrr}
  \hline
 & Horvath & Hannum & PhenoAge & GrimAge\\ 
  \hline
Horvath & 1 &  & &  \\ 
 Hannum & 0.85 & 1 & &  \\ 
 PhenoAge & 0.08 & 0.12 & 1 &  \\ 
 GrimAge & 0.05 & 0.14 & 0.49 & 1 \\
   \hline
\end{tabular}
\end{table}

The estimated correlations are reported in Table \ref{tab:epicorr}, and are close to the empirical correlations computed from the raw data.

\begin{table}[h]
\centering
\caption{Posterior mean and [95\%CI] for the correlations between the epigenetic clocks.}
\begin{tabular}{cc:cc}
\hline
\text{Horvath} & & & \\
0.848 \:[0.807 - 0.882] & \text{Hannum} & & \\
\hdashline
\multicolumn{2}{c}{\multirow{2}{*}{$0.170\: [0.092 - 0.285]$}} & \text{PhenoAge} &\\
 & & 0.446\:[0.371 - 0.572] & \text{GrimAge}\\
 \hline
\end{tabular}
\label{tab:epicorr}
\end{table}

\subsection{Multivariate disease mapping}
\label{sec:motv2}

\begin{figure}[ht!]
    \centering
    \includegraphics[width=0.99\textwidth]{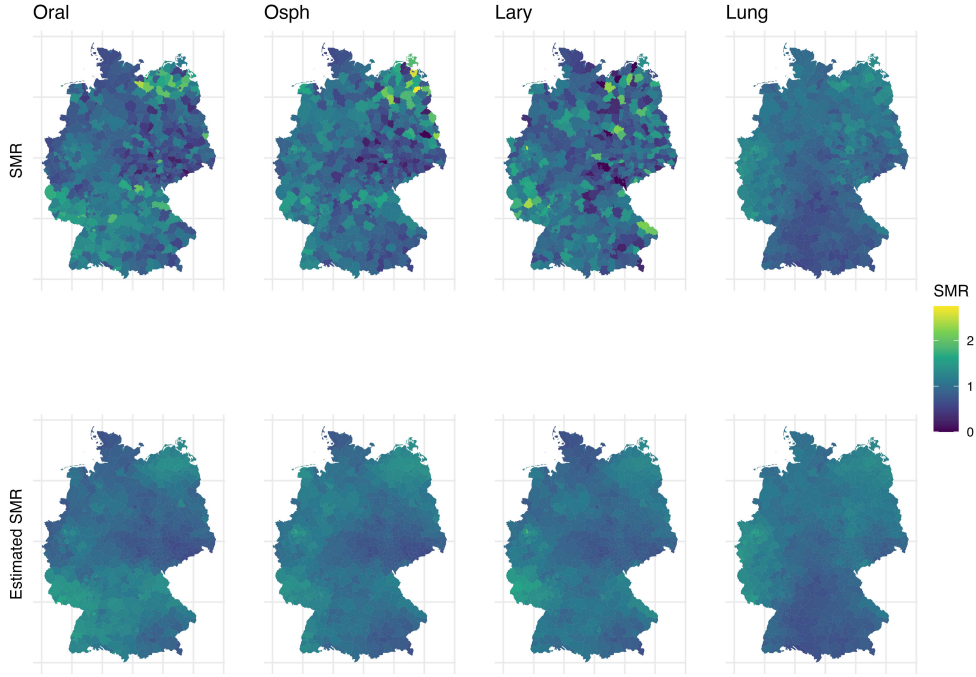}
    \caption{Observed SMR (top maps) and estimated SMR from $M_E$ (bottom maps).}
    \label{fig:smrmaps}
\end{figure}

In our second application, we fit a multivariate disease mapping for lung, oral, oesophagus and larynx cancer men mortality data in 544 districts of Germany from 1986 to 1990 \cite{natario2003non, Held2005} (see figure \ref{fig:smrmaps}, for the Standard Mortality Ratios (SMR) plots). 

The diseases $y_{ij}$ where $j=1,\dots,j$ are distributed as counts over the small area unit $i=1,\dots,544 $. We assume a Poisson distribution with mean $\theta_{ij}= E_{ij} r_{ij}$, where $E_{ij}$ is the expected number of cases (computed based on the demographic characteristics of a reference population) and $r_{ij}$ is the relative risk: 
\[
y_{ij} |E_{ij} r_{ij} \sim \textrm{Pois}(E_{ij} r_{ij}).
\]
The Besag  multivariate disease model  that smooths the risks is modelled as follows:
\begin{equation*}
\log(r_{ij})=\alpha_j + x_{ij}
\end{equation*}
where $\alpha_j$ is the log baseline risk for the $j$-th disease, 
and the term $x_{ij}$ is a random effect to capture 
the variability in $i$ region for $j$-th disease. Thus, let $\mm X$ be the random effect matrix, with $X_{i\mathord{\cdot}}, (i=1\dots,n)$ indicating the $i-th$ row, and $X_{\mathord{\cdot}j}, (j=1,\dots,J)$ the $j-th$ column, the $\text{Vec}({\mm X})=(X_{\mathord{\cdot}1}^T, \cdots, X_{\mathord{\cdot}j}^T)^T$. The $\text{vec\mm(\mm)}$ is defined using a Normal
distribution with zero mean and a highly structured precision matrix $\mm Q$. This representation is a multivariate Gaussian vector, with a sparse precision matrix 
  \begin{equation}
  \label{eq:eqmcar1} 
  \text{Vec}(\mm X) \sim \mathcal{N}(0, \mm Q)
  \end{equation}
where $\mm Q =\mm{R}_{dis} \otimes \mm{R}_{sp}$, are the
$j \times j$ disease inverse correlation and the $n \times n$ spatial correlation matrix respectively, and $\otimes$ indicating the Kronecker product.
For spatial component represented by $\mm R_{sp}$, 
we considered a precision matrix structure \cite{Riebler2016}
defined as: 
\[
\mm R_{sp} = \left\{ \begin{array}{rl}
n_i & \textrm{if } i = g \\
-1 & \textrm{if } i\sim g \textrm{ (area $i$ is neighbour to area $g$)} \\
0 & \textrm{otherwise.}
\end{array}
\right.
\]
That generates a neighbouring graph as report in Figure \ref{fig:figB1}, appendix \ref{sec:appB}.                       
The disease and spatial components of this model are independent, hence for the spatial component we have assumed an ICAR and a BYM2 random effects and specified PC priors on these components, see appendix \mm{B}, for all the details. 

The disease component $\mm{R}_{dis}$ is defined by a graph for the four types of cancer based on previous work in the field \cite{Held2005}. In Table \ref{tab:dmodels}  for model $M_E$, we draw a graph to express that  Oesophagus (Osph), Larynx(Lary), and Oral cancer exhibit higher correlations, indicative of shared risk factors or underlying biological mechanisms. Lung cancer, however, has been observed to have a lower correlation with the other three cancer types. In detail, we assume parent 1 ($p_1$) as the common ancestor to all children (i.e., cancer types), including Lung ($c_4$) as its direct children. Larynx ($c_3$), Oesophagus ($c_2$) and Oral ($c_1$) share an additional common ancestor ($p_2$), making the correlation between them higher compared to their correlation with Lung cancer. Finally, Oral cancer has an additional latent factor ($p_3$), allowing for a lower correlation with Lung cancer compared to the other two cancer types as observed in \cite{Held2005}. We also admit a simplified graph (see Table \ref{tab:dmodels} model $M_E$) that allows Oral cancer ($c_1$) to have the same correlation to Lung cancer as Larynx ($c_3$) and Oesophagus ($c_2$), by removing parent 3. 

\begin{sidewaystable}
\caption{Description of the fitted models.} 
\begin{tabular}{clcclc}
  \hline
Model & Disease & Graph Disease &  Spatial Effect & Model Equations & Parameters \\ 
\hline
$M_A$ & Independent & \includegraphics[scale=0.5]{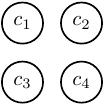} & Correlated (BYM2) &  
$\begin{array}{c} 
y_{j_i}\sim \textrm{Pois}(E_{j_i} r_{j_i})\\
\log(r_{j_i})=\alpha +x_{j_i}\\

\pmb{x}\sim BYM2(\tau_{bym},\phi_{bym}) \\
j=1,\dots,4, i=1,\dots,n\\
\end{array}$ & 8  \\ \hline
$M_B$ & Unstructured correlation & \includegraphics[scale=0.5]{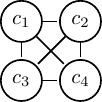} & Independent & 
$\begin{array}{c} 
y_{ij}  \sim \textrm{Pois}(E_{ij} r_{ij})\\
\log(rij_i) =\alpha_j+v_{j}\\
\pmb{v}\sim N(0,\pmb\sigma\pmb C \pmb\sigma^\top)\\ \pmb C \sim LKJ(\eta), \sigma_j \sim \text{gamma}(1,1)\\
\end{array}$ & 10 \\ \hline
$M_C$ & Unstructured correlation & \includegraphics[scale=0.5]{4Co.pdf}& Correlated (Besag) & 
$\begin{array}{c} 
y_{ij} \sim \textrm{Pois}(E_{ij} r_{ij})\\
\log(rij_i) =\alpha_j +v_{j}+x_i\\
\pmb{x}\sim ICAR(\tau_{icar}) \\
\pmb{v}\sim N(0,\pmb\sigma\pmb C \pmb\sigma^\top)\\ \pmb C \sim LKJ(\eta), \sigma_j \sim \text{gamma}(1,1)\\ 
\end{array}$ &10 \\ \hline
$M_D$ & User specified  &\includegraphics[scale=0.5]{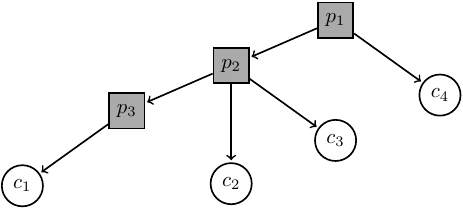} & Correlated (Besag) & 
$\begin{array}{c} 
y_{ij} \sim \textrm{Pois}(E_{ij} r_{ij})\\
\log(rij_i) =\alpha_j +x_i+v_j\\
\pmb{x}\sim ICAR(\tau_{icar}) 
\end{array}$
&7 \\ 
$M_E$ & User specified  &\includegraphics[scale=0.5]{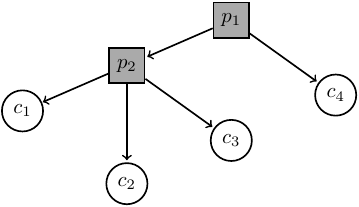} & Correlated (Besag) & 
$\begin{array}{c} 
\pmb{v}\sim N(0,\pmb\sigma\pmb C \pmb\sigma^\top)\\ \pmb C \sim PC_{\text{graph}}, \sigma_j \sim \text{gamma}(1,1)\\ \\ \\ \\
\end{array}$
& 6 \\  \hline
\end{tabular}
\label{tab:dmodels}
\end{sidewaystable}

We fitted five models with various levels of flexibility to illustrate how our new approach offers a trade-off between the model assuming independence and the model with full correlations, both in terms of flexibility and computational burden, see Table \ref{tab:dmodels}. 

First, we fitted the Besag-York-Molli\'e  (BYM) model \cite{Besag1991} 
for each vector in $\mm x$ corresponding to each one of the diseases, independently, as model $M_A$. See appendix \ref{sec:appB} for PC prior for the spatial component.    
In the second model $M_B$, the disease part is modelled with an unstructured correlation matrix, for which we used a PC prior for
the precision parameters and LKJ prior \cite{lewandowski2009generating} for the correlation parameters with $\eta=1$, and an independent identically distributed random effect for the spatial areas. While model $M_C$ is the same, except the spatial part is an MCAR. The last two models, $M_D, M_E$ present a correlation derived from the graph for the disease and an MCAR for the spatial part, respectively. 
These models assume the same spatial smoothing 
for each disease and that the correlation among diseases 
is the same in the different parts of the study area.
Finally, to evaluate and compare the five fitted models, we reported for each the Deviance Information Criteria (DIC), the Widely applicable Bayesian information criterion (WAIC), and the Conditional predictive ordinate using either leave-one-out (CPO) or leave-group-out (GCPO, $n=5$) cross-validation, 
with level set $n=5$, which means that at least $5$ data points are removed to cross-validate each observation, 
see \cite{Liu2022,art720} for details.

Overall, the models considering spatial and between-diseases correlation ($M_C$, $M_D$, $M_E$) performed better 
compared to models assuming independence between diseases ($M_A$) 
or spatial independence ($M_B$). Furthermore, models based on our approach ($M_D$ and $M_E$) performed well compared to the model with the highest flexibility ($M_C$).
On the performance indexes, there are minor differences in the DIC, WAIC, CPO and GCPO 
between $M_D$, $M_E$ and $M_C$ with no clear model outperforming the others. Our approach, hence, manages to provide a fit similar to the most complex model at a lower cost and with a natural interpretation. While the difference in the number of parameters is limited, this difference increases a lot when the number of diseases to model increases, making our approach scalable where the flexible model with unstructured correlation quickly reaches limitations in terms of the computational burden. 

\begin{table}[ht]
\centering
\caption{The DIC, WAIC, CPO and GCPO with the five models fitted.} 
\begin{tabular}{rrrrr}
  \hline
Model  & DIC & WAIC & CPO & GCPO $(n=5)$\\ 
  \hline
MA & 13811.83 & 13736.36 & 7057.73 & 7033.25 \\ 
  MB & 14039.89 & 13932.70 & 7458.40 & 7158.05 \\ 
  MC & \pmb{13686.34} & 13639.37 & \pmb{6945.98} & 6871.62 \\ 
  MD & 13711.37 & 13614.95 & 6968.17 & 6874.21 \\ 
  ME & 13713.65 & \pmb{13608.27} & 6970.68 & \pmb{6864.26} \\ 
   \hline
\end{tabular}

\end{table}
Figure~\ref{fig:smrmaps} displays the four maps of the raw observed and 
estimated SMR based on $M_E$.
As expected, the estimated SMR maps are a smoothed version of the observed ones
highlighting the spatial pattern, which is similar among Oral, Oesophagus and Larynx 
cancer risk and different for Lung cancer.

The estimated correlation between the log risk of each disease considering $M_D$ is shown in Table ~\ref{tab:smrcorr}. The highest correlation is between the Oesophagus and the Larynx, similar to Oral- Oesophagus and Oral-Larynx, with the lowest observed for Oral-Lung, in line with the graph for model $M_D$ in Table \ref{tab:dmodels}.

\begin{table}[ht]
\centering
\caption{Posterior mean and [95\% CI] for the correlations between the logarithm of the relative risk.}
\begin{tabular}{c:ccc}
\hline
 Oral & & & \\
\multirow{2}{*}{0.79 [0.67 - 0.86]}& Oesophagus & &\\
& 0.80 [0.69 - 0.87]& Larynx &\\ 
\hdashline
0.61 [0.49 - 0.70] & \multicolumn{2}{c}{0.61 [0.50 - 0.70]} & Lung \\ 
\hline
\end{tabular}
\label{tab:smrcorr} 
\end{table}

The simplified model $M_E$ assumes only two distinct correlations:
(Oral, Oesophagus), (Oral, Larynx) and (Oesophagus, Larynx) 
share the same correlation of 0.72 (0.60 - 0.81), 
and the correlation between Lung and the others 0.62 (0.52, 0.71). From these values and the goodness-of-fit metrics, one can see that the simplified model returns a similar fit and similar correlations at a lower complexity and with slightly better WAIC and GCPO.

\section{Conclusions}
\label{sec:disc}

Using an intuitive graphical construction, this paper introduced an approach to defining correlation matrices for observed variables or random effects. The essence of our approach is rooted in the interpretation of the model structure, unlike the conventional methods, which offer an unstructured and often ambiguous interpretation of correlations. Each element of the correlation matrix is no longer an isolated parameter but a piece of a coherent, inspired narrative that describes the intricate relationships among variables.

This new approach offers several notable advantages. Firstly, it aligns the statistical modelling process more closely with prior knowledge and empirical observations than covariance models. Researchers and practitioners can map their theoretical constructs and hypotheses directly onto the statistical model, facilitating the union of theory and practical analysis. Hence, this approach enhances the interpretability and transparency of the correlation structure. 
Every estimated correlation is embedded within a framework that makes the results more accessible and interpretable for a broader audience, including those without advanced statistical training. 

Secondly, our approach improves interpretability and eases the computational burden by streamlining the models and decreasing the necessary number of parameters. This introduces a balance between the excessively simplistic independence assumption and the computationally demanding unstructured correlation structures commonly found in conventional practices.

Lastly, the multivariate penalized priors derived are proper by construction, and posterior proprieties are guaranteed. It also provides a stable choice of prior if no other expert choice is available since the prior ensures contraction to a less complex model if the data lacks support for the complex model. 

We have restricted our interpretation to a subclass of directed acyclic graphs to avoid non-suitable correlation matrices. Our starting point was to have a drawing (a graph) based on how researchers hypothesize the multiple variables are related.  
We acknowledge that this is a convenient approach and privileges simple graphs, being aware that other types of graphs could have generated similar correlation matrices, but for which we could not have a better understanding. This method suits a manageable dimension, as we privilege the interpretation. Indeed in \textbf{R-INLA}, it is possible to fit models in high dimensions, but then it's not easy to design the graph structure.

 However, there are limitations. First,  we recognise that the correlation/covariance prior choice remains unsolved even in high-dimensional multivariate response modelling. For example, Chakraborty et al.\cite{ chakraborty2023bayesian} has developed a two-stage modelling approach for multivariate probit model; however, they found that the LKJ prior is too limiting and ended up focusing the inference on the marginals, and bypass a joint prior specification for the marginals. Similarly, Bottolo et al.\cite{bottolo2021computationally} specified an Inverse-Wishart prior in their high-dimensional seemingly unrelated regression modelling. 
 
Second, for the same correlation signs, the user just needs to specify the graph. However, if the user needs to specify different correlation signs, this has to be pre-defined with the graph, as there is no automatic way to construct graphs that allow different signs. 

Lastly, as with any modelling approach, misspecification due to the wrong graph chosen could occur. We have shown in the simulations that correlation estimates can help the user identify potential issues at the graph level (i.e., identical values when the hierarchy is incorrect). 
Hence, different graphs could be fitted to identify the model that at best represents the data-generating process. 

Indeed, future research should focus on developing multivariate penalized complexity before correlation matrices and extending this approach beyond the multivariate regression models, allowing for graphs other than tree graphs or latent graphical models. 

In conclusion, the graphical framework provides interpretable correlation matrices, and the automatic prior construction uses a sequence of simpler models embedded in the penalized complexity prior framework.

\newpage
\begin{appendices}

\section{Computation of the PC prior for a complex exchangeable matrix}
\label{appendix:appA}
\renewcommand{\thefigure}{A\arabic{figure}}
\setcounter{figure}{0}

\begin{figure}[h!]
         \centering
         \begin{tikzpicture}[shorten >=1pt,node distance=2cm,on grid,auto]
    \tikzstyle{state}=[shape=circle,thick,draw,minimum size=0.6cm]
    \tikzstyle{statesq}=[shape=rectangle,thick,draw,minimum size=0.6cm]
    \node[statesq, fill={rgb:black,1;white,2}] (b) {$p_1$};
    \node[state,below left of=b] (Z1) {$c_1$};
     \node[state,below of=b](Z2)  {$c_2$};
    \node[state,below right of=b] (Z3) {$c_3$};
       \path[->,draw,thick]
    (b) edge node {} (Z1)
    (b) edge node {} (Z2)
    (b) edge node {} (Z3);
     \end{tikzpicture} 
        \caption{One parameter graph with two variables.  Grey and white nodes represent latent factors and observed variables, respectively.}
        \label{fig:fig1}
\end{figure}
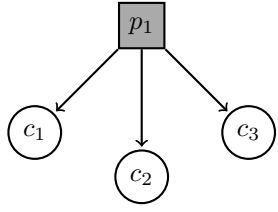

We derive first the exact PC prior for an exchangeable correlation matrix (one latent factor and 2 children, see Figure \ref{fig:fig1}) correlation matrix is:

\begin{equation*}
    \label{eq:C1}%
    \mm{C}[1:3,1:3]=\begin{bmatrix}{}
        1    &  \frac{q_1^2}{1+q_1^2}& \frac{q_1^2}{1+q_1^2} \\ 
      \frac{q_1^2}{1+q_1^2}  & 1  & \frac{q_1^2}{1+q_1^2 } \\
      \frac{q_1^2}{1+q_1^2}   &   \frac{q_1^2}{1+q_1^2 } & 1 \\ 
          \end{bmatrix}
        \end{equation*}

and the simpler model with an identity matrix $\mm C_0= \mm I$. 
The KLD is: 

\[
\text{KLD}= \frac{1}{2} \left\{\text{tr}(\mm C_0^{-1}\mm C_1) - p - \log\left(\frac{|\mm C_1|}{|\mm C_0|}\right)\right\}
\]
where $p=2$ (dimension of $\mm C_1$), $\text{tr}(\mm C_0^{-1}\mm C_1)=2$, $|\mm C_0|=1$ and $|\mm C_1|=\frac{2q_1^2+1}{(1+q_1^2)^2}$,
\[ 
\text{KLD} =\frac{1}{2}\left(2 - 2 - \log\left(\frac{2q_1^2+1}{(1+q_1^2)^2}\right)\right).
\]
Then, the distance is reduced to:  
\[
\text{Dist} = \sqrt{2 \cdot \text{KLD}} =\sqrt{-\log\left(\frac{2q_1^2+1}{(1+q_1^2)^2}\right)}
\]
Hence, the prior for $\theta=\log(q_1^2)$ is:

\[ 
\pi(\theta) = \lambda \exp\left(-\lambda \cdot \sqrt{-\log\left(\frac{2q_1^2+1}{(1+q_1^2)^2}\right)}\right) \left|\frac{\partial \text{Dist}}{\partial \theta}\right|
\]  
Using the second order expansion instead, we get $\text{Dist}= q_1^2$, so that $\pi(\theta) = \lambda \exp(-\lambda \exp(\theta)) \exp(\theta)$.

\newpage
\section{Penalized Complexity Prior for ICAR and BYM2 }
\label{sec:appB}
\renewcommand{\thefigure}{B\arabic{figure}}
\setcounter{figure}{0}
\subsubsection*{Spatial graph}

The neighbouring structure also can be represented by a graph, where every node is a district area
\begin{figure}[h!]
\centering
\includegraphics[scale=0.8]{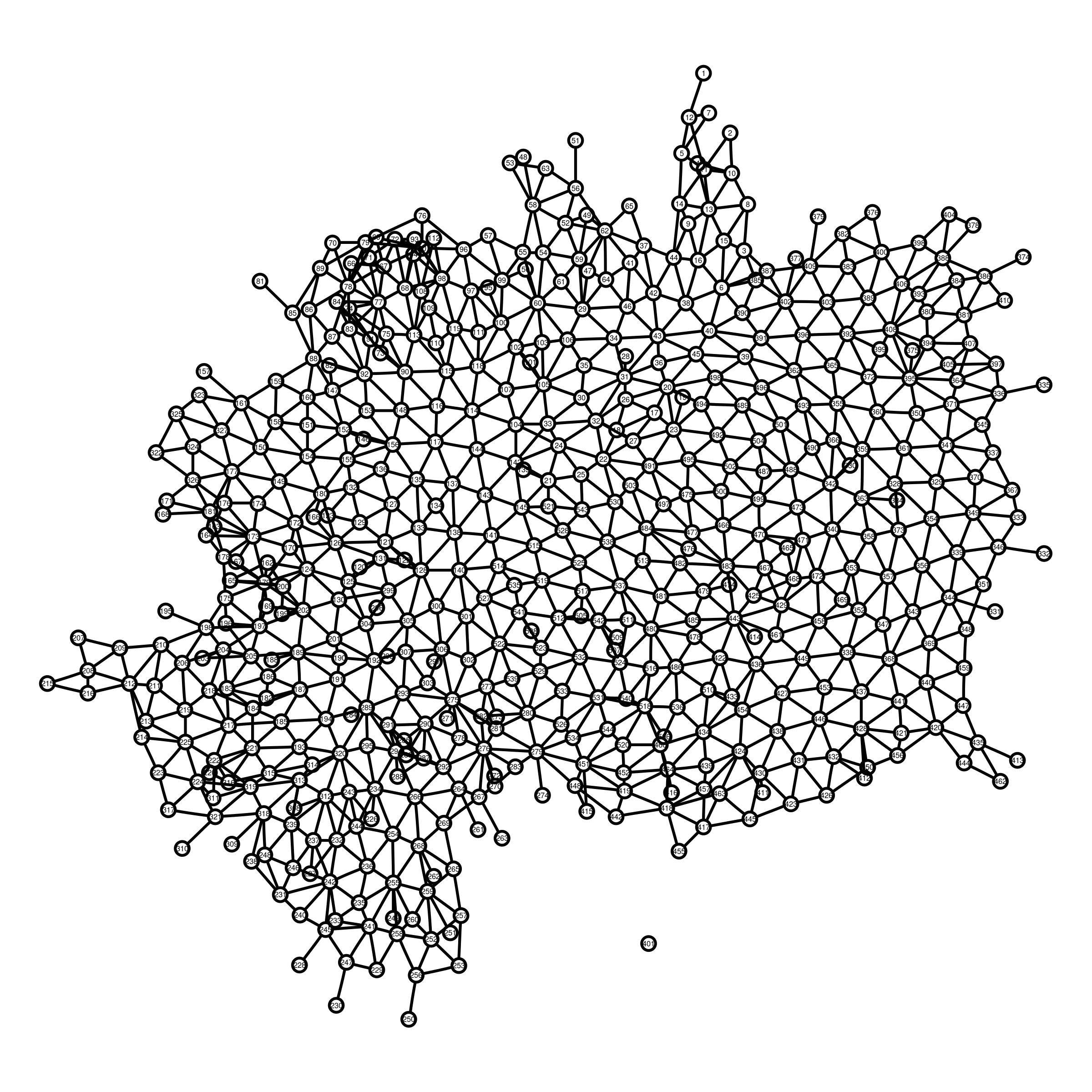}
\caption{The graph induced by the Germany district map.}
\label{fig:figB1}
\end{figure}

\subsubsection*{PC Priors}

For the spatial component in the multivariate disease mapping, we have assumed an intrinsic conditional autoregressive (ICAR) component for spatial smoothing, for  model  $M_C$, $M_D$ and $M_E$ with ICAR. 

Given a set of observations taken at $n$ different areal units of a region, spatial interactions between a pair of units $n_i$ and $n_{i^{\prime}}$ can be modelled conditionally as a spatial random variable $\mm{b}$, which is an n-length vector $\mm{b}=(b_1,\dots,b_n)$: 
\[
b_i|\pmb{b}_{-i},\tau_b \sim \mathcal{N}(\frac{1}{n_{\delta_i}}\sum_{j \in \delta_i} b_j, \frac{1}{n_{\delta_j}\tau_b})
\]
where $\tau_b$ is a precision parameter. The mean of $b_i$ equals the mean of the effects over all neighbours, and the precision is proportional to the number of
neighbours. The joint distribution for $\pmb{b}$ is:
\[
\pi(\pmb{b}|\tau_b) \propto \exp(-\frac{\tau_b}{2}\pmb{b^TR_{sp}b}).
\]

For model $M_A$, we opted for a re-parametrization of ICAR, namely called BYM2. In this case,  the spatial effect is split into two component $\pmb{b=u+v}$, one  to account for the a structured spatial effect and an unstructured ordinary random-effects component for non-spatial heterogeneity. 
There are many parameterization possible for this, we have used a re-parameterization by Reiebler et al. \cite{Riebler2016} where the spatial random effect, is assumed to follow a normal distribution with mean zero and covariance matrix:
 \[
var(\pmb{b}|\tau_b, \phi)=\tau_b^{-1}((1-\phi)\pmb{I}+ \phi\pmb{Q})^{-1}
 \]
Where the $\phi$ is the mixing parameter that 'splits' the variance between a structured and unstructured spatial component.

The PC prior for the $\tau_{b}$ is derided by the comparisons between the base model $\Sigma_{base}=0$ and a complex model $\Sigma_{flex}=var(\pmb{b}|\tau_b, \phi)$
\[
\pi(\tau_b)=\frac{\theta}{2}\tau_b^{-3/2}exp(-\theta\tau_b^{-1/2}) .
\]

The PC prior for $\phi$  compares $\Sigma_{base}=\pmb{I}$ and $\Sigma_{flex}(\phi)=(1-\phi)\pmb{I}+\phi\pmb{Q}_*^-$.
A reasonable formulation might be $Prob(\phi < 0.5) = 2/3$, which gives more density mass to values of $\phi$ smaller than 0.5. 
Because the  parameter $\phi$ represents the fraction of the total variance which can be attributed to spatial dependency
structure, this translate in a conservative approach where the unstructured random effects accounts more of the variability than the spatially structured effect \cite{Riebler2016}. 

\newpage 
\section{Simulations results for low correlations}
\label{sec:appC}

\begin{table}[!ht]
\caption{Simulations results with 3 parents and 4 children for $\lambda=1, 3$ and $5$. The column ``misspecified'' corresponds to a model with the wrong hierarchical order for the correlation structure of children (using $\lambda=3$), and the column ``unspecified'' corresponds to the unspecified correlation between all children (results presented as mean [95\% CI]). Computation time is provided in seconds in the last row.}
\footnotesize
{\tabcolsep=2pt \renewcommand{\arraystretch}{1.5}
\begin{tabular}{@{}lccccccccccccccccccccccccccccc@{}}
\hline
True values & $\lambda$ = 1 & $\lambda$ = 3 & $\lambda$ = 5 & misspecified & unspecified\\
\hline
\multicolumn{6}{c}{N=30 individuals}\\\hline
$\sigma_{c1}$=1 & 1.02 [0.8, 1.25] & 1 [0.79, 1.23] & 0.99 [0.78, 1.21] & 1.01 [0.79, 1.24] & 0.99 [0.78, 1.21] & \\
 $\sigma_{c2}$=0.2 & 0.2 [0.16, 0.25] & 0.2 [0.16, 0.24] & 0.2 [0.15, 0.24] & 0.2 [0.16, 0.25] & 0.27 [0.24, 0.3] & \\
 $\sigma_{c3}$=0.1 & 0.1 [0.07, 0.13] & 0.1 [0.07, 0.12] & 0.1 [0.07, 0.12] & 0.1 [0.07, 0.12] & 0.21 [0.19, 0.22] & \\
 $\sigma_{c4}$=0.5 & 0.51 [0.4, 0.62] & 0.5 [0.39, 0.61] & 0.5 [0.39, 0.6] & 0.51 [0.39, 0.62] & 0.52 [0.42, 0.63] & \\
 $\rho_1$=0.4 & 0.45 [0.21, 0.67] & 0.4 [0.18, 0.63] & 0.35 [0.15, 0.58] & 0.18 [0.06, 0.37] & 0.3 [0.03, 0.52] & \\
 $\rho_2$=0.4 & 0.45 [0.26, 0.65] & 0.39 [0.22, 0.6] & 0.35 [0.18, 0.54] & 0.18 [0.06, 0.37] & 0.18 [0.06, 0.31] & \\
 $\rho_3$=0.2 & 0.18 [0.05, 0.38] & 0.16 [0.04, 0.34] & 0.14 [0.04, 0.3] & 0.18 [0.06, 0.37] & 0.12 [-0.04, 0.26] & \\
 \midrule $\beta_{01}$=0.2 & 0.19 [-0.17, 0.51] & 0.19 [-0.17, 0.51] & 0.19 [-0.17, 0.51] & 0.19 [-0.17, 0.51] & 0.19 [-0.17, 0.52] & \\
 $\beta_{11}$=-0.1 & -0.1 [-0.19, -0.02] & -0.1 [-0.19, -0.02] & -0.1 [-0.19, -0.02] & -0.1 [-0.19, -0.02] & -0.1 [-0.19, -0.02] & \\
 $\beta_{02}$=0.2 & 0.2 [0.17, 0.23] & 0.2 [0.17, 0.23] & 0.2 [0.17, 0.23] & 0.2 [0.17, 0.23] & 0.2 [0.17, 0.23] & \\
 $\beta_{12}$=-0.1 & -0.1 [-0.27, 0.07] & -0.1 [-0.27, 0.07] & -0.1 [-0.27, 0.07] & -0.1 [-0.27, 0.07] & -0.1 [-0.27, 0.07] & \\
 \midrule
 Comp. time (s) & 0.94 [0.82, 1.19] & 0.93 [0.83, 1.22] & 0.96 [0.81, 1.37] & 0.97 [0.82, 1.27] & 1.63 [1.47, 2.91] & \\
\hline
\multicolumn{6}{c}{N=1000 individuals}\\ \hline
$\sigma_{c1}$=1 & 1 [0.94, 1.04] & 1 [0.94, 1.04] & 1 [0.94, 1.04] & 1 [0.94, 1.04] & 1 [0.94, 1.04] & \\
 $\sigma_{c2}$=0.2 & 0.2 [0.19, 0.21] & 0.2 [0.19, 0.21] & 0.2 [0.19, 0.21] & 0.2 [0.19, 0.21] & 0.2 [0.2, 0.21] & \\
 $\sigma_{c3}$=0.1 & 0.1 [0.1, 0.1] & 0.1 [0.1, 0.1] & 0.1 [0.1, 0.1] & 0.1 [0.1, 0.1] & 0.1 [0.1, 0.11] & \\
 $\sigma_{c4}$=0.5 & 0.5 [0.48, 0.52] & 0.5 [0.48, 0.52] & 0.5 [0.48, 0.52] & 0.5 [0.48, 0.52] & 0.5 [0.48, 0.52] & \\
 $\rho_1$=0.4 & 0.4 [0.35, 0.44] & 0.39 [0.35, 0.43] & 0.39 [0.34, 0.43] & 0.26 [0.23, 0.29] & 0.39 [0.34, 0.43] & \\
 $\rho_2$=0.4 & 0.4 [0.35, 0.44] & 0.39 [0.35, 0.44] & 0.39 [0.34, 0.44] & 0.26 [0.23, 0.29] & 0.38 [0.34, 0.42] & \\
 $\rho_3$=0.2 & 0.2 [0.16, 0.23] & 0.2 [0.16, 0.23] & 0.19 [0.16, 0.23] & 0.26 [0.23, 0.29] & 0.19 [0.16, 0.23] & \\
 \midrule $\beta_{01}$=0.2 & 0.2 [0.14, 0.26] & 0.2 [0.14, 0.26] & 0.2 [0.14, 0.26] & 0.2 [0.14, 0.26] & 0.2 [0.14, 0.26] & \\
 $\beta_{11}$=-0.1 & -0.1 [-0.11, -0.09] & -0.1 [-0.11, -0.09] & -0.1 [-0.11, -0.09] & -0.1 [-0.11, -0.09] & -0.1 [-0.11, -0.09] & \\
 $\beta_{02}$=0.2 & 0.2 [0.19, 0.21] & 0.2 [0.19, 0.21] & 0.2 [0.19, 0.21] & 0.2 [0.19, 0.21] & 0.2 [0.19, 0.21] & \\
 $\beta_{12}$=-0.1 & -0.1 [-0.13, -0.07] & -0.1 [-0.13, -0.07] & -0.1 [-0.13, -0.07] & -0.1 [-0.13, -0.07] & -0.1 [-0.13, -0.07] & \\
 \midrule 
 Comp. time (s) & 17.53 [15.98, 20.91] & 17.59 [15.56, 20.6] & 17.74 [15.83, 20.71] & 23.64 [17.41, 38.68] & 20.28 [18.96, 22.47]\\
\hline
\end{tabular}}
\label{tab:APPEx3p4c}
\end{table}
\clearpage

\begin{table}[!ht]
\caption{Simulations results with 7 parents and 8 children for $\lambda=1, 3$ and $5$ with 30 individuals per dataset. The column ``unspecified'' corresponds to the unspecified correlation between all children (results presented as: mean [95\% CI]). Computation time is provided in seconds in the last row.}
\footnotesize
\centering
{\tabcolsep=2pt \renewcommand{\arraystretch}{1.5}
\begin{tabular}{@{}lccccccccccccccccccccccccccccc@{}}
\hline
\multicolumn{5}{c}{N=30 individuals}\\
True values & $\lambda$ = 1 & $\lambda$ = 3 & $\lambda$ = 5 & unspecified \\
\hline
$\sigma_{c1}$=1 & 1 [0.73, 1.3] & 0.97 [0.72, 1.26] & 0.96 [0.7, 1.24] & 0.95 [0.71, 1.22] & \\
 $\sigma_{c2}$=0.5 & 0.5 [0.39, 0.67] & 0.49 [0.38, 0.65] & 0.49 [0.38, 0.64] & 0.5 [0.41, 0.64] & \\
 $\sigma_{c3}$=1 & 1.02 [0.75, 1.3] & 1 [0.73, 1.28] & 0.98 [0.72, 1.26] & 0.96 [0.73, 1.24] & \\
 $\sigma_{c4}$=0.5 & 0.51 [0.42, 0.62] & 0.5 [0.41, 0.6] & 0.5 [0.4, 0.59] & 0.52 [0.42, 0.61] & \\
 $\sigma_{c5}$=1 & 1.05 [0.8, 1.3] & 1.02 [0.79, 1.26] & 1.01 [0.78, 1.24] & 1 [0.77, 1.23] & \\
 $\sigma_{c6}$=0.5 & 0.52 [0.39, 0.63] & 0.5 [0.39, 0.62] & 0.5 [0.38, 0.61] & 0.51 [0.4, 0.63] & \\
 $\sigma_{c7}$=1 & 1.01 [0.78, 1.23] & 0.99 [0.77, 1.2] & 0.97 [0.76, 1.18] & 0.97 [0.73, 1.3] & \\
 $\sigma_{c8}$=0.5 & 0.5 [0.36, 0.63] & 0.49 [0.35, 0.61] & 0.49 [0.35, 0.6] & 0.51 [0.38, 0.63] & \\
 $\rho_1$=0.4 & 0.48 [0.28, 0.68] & 0.43 [0.25, 0.63] & 0.38 [0.21, 0.58] & 0.38 [0.02, 0.67] & \\
 $\rho_2$=0.4 & 0.48 [0.29, 0.68] & 0.42 [0.25, 0.62] & 0.37 [0.21, 0.57] & 0.37 [0.04, 0.64] & \\
 $\rho_3$=0.4 & 0.48 [0.3, 0.64] & 0.42 [0.25, 0.59] & 0.37 [0.22, 0.53] & 0.39 [0.06, 0.61] & \\
 $\rho_4$=0.4 & 0.48 [0.3, 0.67] & 0.42 [0.26, 0.6] & 0.37 [0.22, 0.54] & 0.36 [0.05, 0.74] & \\
 $\rho_5$=0.3 & 0.3 [0.14, 0.51] & 0.26 [0.12, 0.46] & 0.23 [0.1, 0.43] & 0.29 [0.07, 0.51] & \\
 $\rho_6$=0.3 & 0.29 [0.12, 0.47] & 0.25 [0.11, 0.42] & 0.22 [0.1, 0.38] & 0.29 [0.05, 0.49] & \\
 $\rho_7$=0.2 & 0.16 [0.04, 0.33] & 0.14 [0.04, 0.29] & 0.12 [0.03, 0.25] & 0.19 [-0.06, 0.4] & \\
 $\rho_8$=0.2 & 0.16 [0.04, 0.33] & 0.14 [0.04, 0.28] & 0.12 [0.03, 0.25] & 0.18 [-0.03, 0.4] & \\
 $\rho_9$=0.2 & 0.16 [0.04, 0.32] & 0.14 [0.03, 0.28] & 0.12 [0.03, 0.25] & 0.19 [-0.05, 0.41] & \\
 $\rho_{10}$=0.2 & 0.16 [0.04, 0.32] & 0.14 [0.04, 0.28] & 0.12 [0.03, 0.25] & 0.18 [-0.04, 0.44] & \\
 \midrule $\beta_{01}$=0.2 & 0.12 [-0.61, 0.86] & 0.12 [-0.6, 0.88] & 0.12 [-0.6, 0.91] & 0.11 [-0.71, 0.96] & \\
 $\beta_{11}$=-0.1 & -0.11 [-0.27, 0.08] & -0.11 [-0.27, 0.08] & -0.11 [-0.27, 0.08] & -0.11 [-0.27, 0.08] & \\
 $\beta_{21}$=-0.2 & -0.2 [-0.82, 0.36] & -0.2 [-0.8, 0.38] & -0.2 [-0.79, 0.4] & -0.18 [-0.82, 0.39] & \\
 $\beta_{31}$=0.1 & 0.16 [-0.44, 0.78] & 0.15 [-0.47, 0.77] & 0.15 [-0.45, 0.77] & 0.15 [-0.69, 0.92] & \\
 $\beta_{02}$=0.2 & 0.17 [-0.58, 0.99] & 0.17 [-0.57, 0.99] & 0.17 [-0.57, 1] & 0.17 [-0.58, 0.93] & \\
 $\beta_{12}$=-0.1 & -0.11 [-0.29, 0.04] & -0.11 [-0.29, 0.04] & -0.11 [-0.29, 0.04] & -0.11 [-0.3, 0.04] & \\
 $\beta_{22}$=-0.2 & -0.18 [-0.89, 0.49] & -0.18 [-0.9, 0.48] & -0.18 [-0.91, 0.48] & -0.15 [-0.87, 0.57] & \\
 $\beta_{32}$=0.1 & 0.12 [-0.48, 0.82] & 0.12 [-0.49, 0.8] & 0.11 [-0.51, 0.8] & 0.11 [-0.58, 0.77] & \\
 $\beta_{03}$=0.2 & 0.19 [-0.75, 1.03] & 0.2 [-0.75, 1.01] & 0.2 [-0.76, 1.01] & 0.18 [-0.75, 0.96] & \\
 $\beta_{13}$=-0.1 & -0.1 [-0.25, 0.05] & -0.1 [-0.25, 0.05] & -0.1 [-0.25, 0.05] & -0.09 [-0.25, 0.05] & \\
 $\beta_{23}$=-0.2 & -0.21 [-0.75, 0.48] & -0.21 [-0.75, 0.46] & -0.2 [-0.75, 0.45] & -0.18 [-0.74, 0.56] & \\
 $\beta_{33}$=0.1 & 0.11 [-0.59, 0.84] & 0.11 [-0.58, 0.81] & 0.1 [-0.57, 0.8] & 0.11 [-0.57, 0.8] & \\
 $\beta_{04}$=0.2 & 0.22 [-0.62, 1.23] & 0.22 [-0.63, 1.21] & 0.22 [-0.6, 1.21] & 0.21 [-0.59, 1.2] & \\
 $\beta_{14}$=-0.1 & -0.11 [-0.27, 0.09] & -0.11 [-0.27, 0.09] & -0.11 [-0.27, 0.09] & -0.1 [-0.27, 0.09] & \\
 $\beta_{24}$=-0.2 & -0.22 [-0.88, 0.44] & -0.22 [-0.89, 0.43] & -0.22 [-0.9, 0.42] & -0.24 [-0.93, 0.43] & \\
 $\beta_{34}$=0.1 & 0.1 [-0.63, 0.78] & 0.1 [-0.64, 0.75] & 0.1 [-0.64, 0.73] & 0.12 [-0.69, 0.85] & \\
 \midrule
 Comp. time (s) & 2.67 [2.23, 3.01] & 2.72 [2.25, 3.15] & 2.75 [2.19, 4.71] & 25.8 [21.25, 30.4] & \\
 \hline
\end{tabular}}
\label{tab:APPEx7p8c}
\end{table}
\clearpage

\begin{table}[!ht]
\caption{Simulations results with 7 parents and 8 children for $\lambda=1, 3$ and 5 with 1000 individuals per dataset. The column ``unspecified'' corresponds to the unspecified correlation between all children (results presented as mean [95\% CI]). Computation time is provided in seconds in the last row.}
\footnotesize
\centering
{\tabcolsep=2pt \renewcommand{\arraystretch}{1.5}
\begin{tabular}{@{}lccccccccccccccccccccccccccccc@{}}
\hline
 \multicolumn{5}{c}{N=1000 individuals}\\
True values & $\lambda$ = 1 & $\lambda$ = 3 & $\lambda$ = 5 & unspecified \\
\hline
$\sigma_{c1}$=1 & 1 [0.96, 1.04] & 1 [0.96, 1.04] & 1 [0.96, 1.04] & 1 [0.96, 1.04] & \\
 $\sigma_{c2}$=0.5 & 0.5 [0.47, 0.53] & 0.5 [0.47, 0.53] & 0.5 [0.47, 0.53] & 0.5 [0.48, 0.52] & \\
 $\sigma_{c3}$=1 & 1 [0.95, 1.04] & 1 [0.96, 1.05] & 1 [0.95, 1.04] & 0.99 [0.96, 1.04] & \\
 $\sigma_{c4}$=0.5 & 0.5 [0.48, 0.52] & 0.5 [0.48, 0.52] & 0.5 [0.48, 0.52] & 0.5 [0.48, 0.52] & \\
 $\sigma_{c5}$=1 & 1 [0.95, 1.04] & 1 [0.95, 1.04] & 0.99 [0.95, 1.04] & 1 [0.95, 1.04] & \\
 $\sigma_{c6}$=0.5 & 0.5 [0.48, 0.52] & 0.5 [0.48, 0.52] & 0.5 [0.48, 0.52] & 0.5 [0.48, 0.52] & \\
 $\sigma_{c7}$=1 & 1 [0.96, 1.05] & 1 [0.95, 1.04] & 0.99 [0.95, 1.04] & 0.99 [0.95, 1.04] & \\
 $\sigma_{c8}$=0.5 & 0.5 [0.48, 0.53] & 0.5 [0.48, 0.53] & 0.5 [0.48, 0.53] & 0.5 [0.48, 0.52] & \\
 $\rho_1$=0.4 & 0.4 [0.36, 0.45] & 0.4 [0.34, 0.44] & 0.4 [0.33, 0.45] & 0.4 [0.35, 0.44] & \\
 $\rho_2$=0.4 & 0.4 [0.34, 0.44] & 0.4 [0.34, 0.44] & 0.39 [0.34, 0.44] & 0.39 [0.33, 0.44] & \\
 $\rho_3$=0.4 & 0.4 [0.35, 0.45] & 0.39 [0.34, 0.45] & 0.39 [0.34, 0.45] & 0.4 [0.36, 0.46] & \\
 $\rho_4$=0.4 & 0.4 [0.34, 0.44] & 0.4 [0.34, 0.44] & 0.4 [0.35, 0.43] & 0.39 [0.34, 0.44] & \\
 $\rho_5$=0.3 & 0.3 [0.26, 0.34] & 0.3 [0.26, 0.33] & 0.3 [0.26, 0.34] & 0.3 [0.26, 0.34] & \\
 $\rho_6$=0.3 & 0.3 [0.26, 0.33] & 0.29 [0.26, 0.33] & 0.29 [0.26, 0.33] & 0.3 [0.26, 0.33] & \\
 $\rho_7$=0.2 & 0.2 [0.17, 0.23] & 0.19 [0.17, 0.23] & 0.19 [0.16, 0.23] & 0.2 [0.16, 0.24] & \\
 $\rho_8$=0.2 & 0.2 [0.17, 0.24] & 0.19 [0.17, 0.23] & 0.19 [0.17, 0.23] & 0.2 [0.16, 0.24] & \\
 $\rho_9$=0.2 & 0.2 [0.17, 0.23] & 0.19 [0.17, 0.23] & 0.2 [0.17, 0.23] & 0.19 [0.15, 0.25] & \\
 $\rho_{10}$=0.2 & 0.2 [0.17, 0.24] & 0.19 [0.16, 0.24] & 0.19 [0.17, 0.23] & 0.19 [0.16, 0.24] & \\
 \midrule $\beta_{01}$=0.2 & 0.2 [0.07, 0.31] & 0.2 [0.07, 0.31] & 0.2 [0.07, 0.31] & 0.19 [0.06, 0.29] & \\
 $\beta_{11}$=-0.1 & -0.1 [-0.13, -0.07] & -0.1 [-0.13, -0.07] & -0.1 [-0.13, -0.07] & -0.1 [-0.12, -0.08] & \\
 $\beta_{21}$=-0.2 & -0.2 [-0.3, -0.09] & -0.2 [-0.3, -0.09] & -0.2 [-0.3, -0.09] & -0.2 [-0.28, -0.1] & \\
 $\beta_{31}$=0.1 & 0.1 [0, 0.2] & 0.1 [0, 0.2] & 0.1 [0, 0.2] & 0.1 [0.01, 0.2] & \\
 $\beta_{02}$=0.2 & 0.19 [0.07, 0.32] & 0.19 [0.07, 0.32] & 0.19 [0.07, 0.32] & 0.19 [0.06, 0.3] & \\
 $\beta_{12}$=-0.1 & -0.1 [-0.13, -0.07] & -0.1 [-0.13, -0.07] & -0.1 [-0.13, -0.07] & -0.1 [-0.12, -0.07] & \\
 $\beta_{22}$=-0.2 & -0.19 [-0.29, -0.1] & -0.19 [-0.29, -0.1] & -0.19 [-0.29, -0.1] & -0.19 [-0.29, -0.09] & \\
 $\beta_{32}$=0.1 & 0.1 [0, 0.2] & 0.1 [0, 0.2] & 0.1 [0, 0.2] & 0.11 [0.01, 0.2] & \\
 $\beta_{03}$=0.2 & 0.2 [0.08, 0.33] & 0.2 [0.08, 0.33] & 0.2 [0.08, 0.33] & 0.18 [0.05, 0.32] & \\
 $\beta_{13}$=-0.1 & -0.1 [-0.13, -0.08] & -0.1 [-0.13, -0.08] & -0.1 [-0.13, -0.08] & -0.1 [-0.12, -0.08] & \\
 $\beta_{23}$=-0.2 & -0.2 [-0.3, -0.09] & -0.2 [-0.3, -0.09] & -0.2 [-0.3, -0.09] & -0.2 [-0.3, -0.1] & \\
 $\beta_{33}$=0.1 & 0.1 [-0.01, 0.19] & 0.1 [-0.01, 0.19] & 0.1 [-0.01, 0.19] & 0.11 [-0.02, 0.19] & \\
 $\beta_{04}$=0.2 & 0.2 [0.08, 0.35] & 0.2 [0.08, 0.35] & 0.2 [0.08, 0.35] & 0.21 [0.08, 0.38] & \\
 $\beta_{14}$=-0.1 & -0.1 [-0.13, -0.07] & -0.1 [-0.13, -0.07] & -0.1 [-0.13, -0.07] & -0.1 [-0.12, -0.07] & \\
 $\beta_{24}$=-0.2 & -0.2 [-0.31, -0.08] & -0.2 [-0.31, -0.08] & -0.2 [-0.31, -0.08] & -0.21 [-0.32, -0.1] & \\
 $\beta_{34}$=0.1 & 0.1 [0, 0.2] & 0.1 [0, 0.2] & 0.1 [0, 0.2] & 0.09 [-0.03, 0.2] & \\
 \midrule
 Comp. time (s) & 123.44 [101.5, 188.46] & 126.14 [101.07, 190.47] & 126.4 [99.72, 196.26] & 471.15 [409.41, 527.3] & \\
 \hline
\end{tabular}}
\label{tab:APPEx7p8c2}
\end{table}
\clearpage

\end{appendices}


\section*{Code and data}
Data and code for this article can be found in \begin{verbatim}
https://github.com/alan-turing-institute/Graphical_CorrelationModels
\end{verbatim} 
\newpage

\bibliography{mybib}

\end{document}